\documentclass{aa}
\usepackage{graphicx}
\usepackage[varg]{txfonts}
\usepackage{units}
\usepackage{mathtools}
\usepackage{color}
\usepackage{natbib,twoopt}
\usepackage[breaklinks=true]{hyperref} 
\bibpunct{(}{)}{;}{a}{}{,}             
\makeatletter
  \newcommandtwoopt{\citeads}[3][][]{\href{http://adsabs.harvard.edu/abs/#3}%
    {\def\hyper@linkstart##1##2{}%
     \let\hyper@linkend\@empty\citealp[#1][#2]{#3}}}
  \newcommandtwoopt{\citepads}[3][][]{\href{http://adsabs.harvard.edu/abs/#3}%
    {\def\hyper@linkstart##1##2{}%
     \let\hyper@linkend\@empty\citep[#1][#2]{#3}}}
  \newcommandtwoopt{\citetads}[3][][]{\href{http://adsabs.harvard.edu/abs/#3}%
    {\def\hyper@linkstart##1##2{}%
     \let\hyper@linkend\@empty\citet[#1][#2]{#3}}}
  \newcommandtwoopt{\citeyearads}[3][][]%
    {\href{http://adsabs.harvard.edu/abs/#3}
    {\def\hyper@linkstart##1##2{}%
     \let\hyper@linkend\@empty\citeyear[#1][#2]{#3}}}
  \renewcommand*\aa@pageof{, page \thepage{} of \pageref*{LastPage}}
\makeatother
\hypersetup{pdfpagemode = {UseNone},
            pdftitle = {Particle accretion onto planets in discs with hydrodynamic turbulence},
            pdfcreator = {\LaTeX},
            pdfauthor = {Giovanni Picogna, Moritz Stoll, Wilhelm Kley},            pdfsubject = {},
            pdfview = {FitH},
            pdfstartview = {FitH},
            colorlinks = {true},
            linkcolor = [rgb]{0,0.35,0.7},
            citecolor = [rgb]{0,0.35,0.7},
            filecolor = [rgb]{0.61,0,0},
            urlcolor = [rgb]{0,0.35,0.7},
           }

\usepackage{afterpage}
\usepackage{silence}
\newcommand\Rey{\mbox{\textit{Re}}}
\newcommand\Sch{\mbox{\textit{Sc}}}
\newcommand\St{\mbox{\textit{St}}}
\newcommand\taus{{\tau_\mathrm{s}}}

\newcommand{\beq}{\begin{equation}}
\newcommand{\eeq}{\end{equation}}

\begin{document}

\title{Particle accretion onto planets in discs with hydrodynamic turbulence}

\author{ Giovanni Picogna\inst{1,2} \and
Moritz~H.~R. Stoll\inst{2} \and
Wilhelm Kley\inst{2}}

\institute{
Universit\"{a}ts-Sternwarte, Ludwig-Maximilians-Universit\"{a}t M\"{u}nchen,
Scheinerstr. 1, D-81679 M\"{u}nchen, Germany\\
\email{picogna@usm.lmu.de}\\
\and
Institut f\"{u}r Astronomie und Astrophysik, Universit\"{a}t T\"{u}bingen,
Auf der Morgenstelle 10, D-72076 T\"{u}bingen, Germany\\
\email{\{moritz.stoll@, wilhelm.kley@\}uni-tuebingen.de}\\
}

\date{Received \today / Accepted \today}

\abstract{
The growth process of protoplanets can be sped up by accreting a large number of solid, pebble-sized objects that are still present in the protoplanetary disc.
It is still an open question on how efficient this process works in realistic turbulent discs.

We investigate the accretion of pebbles in turbulent discs that are driven by the purely hydrodynamical vertical shear instability (VSI).
For this purpose, we performed global 3D simulations of locally isothermal, VSI turbulent discs that have embedded protoplanetary cores from $5$ to \unit[$100$]{M$_\oplus$}, which are placed at \unit[$5.2$]{au} distance from the star.
In addition, we followed the evolution of a swarm of embedded pebbles of different sizes under the action of drag forces between gas and particles in this turbulent flow.
Simultaneously, we performed a set of comparison simulations for laminar viscous discs where the particles experience stochastic kicks.
For both cases, we measured the accretion rate onto the cores as a function of core mass and Stokes number ($\tau_\mathrm{s}$) of the particles and compared these values to recent magneto-rotational instability (MRI) turbulence simulations.

Overall the dynamic is very similar for the particles in the VSI turbulent disc and the laminar case with stochastic kicks.
For small mass planets (i.e. \unit[$5-10$]{M$_\oplus$}), well-coupled particles with $\tau_\mathrm{s} = 1$, which have a size of about one metre at this location, we find an accretion efficiency (rate of particles accreted over drifting inwards) of about $1.6-3\%$.
For smaller and larger particles this efficiency is higher.
However, the fast inwards drift for $\tau_\mathrm{s}=1$ particles makes them the most effective for rapid growth, leading to mass doubling times of about \unit[$20,000$]{yr}.
For masses between $10$ and \unit[$30$]{M$_\oplus$} the core reaches the pebble isolation mass and the particles are trapped at the pressure maximum just outside of the planet, shutting off further particle accretion.
}


\maketitle

\titlerunning{Particle accretion in VSI turbulent discs}
\authorrunning{Picogna et al.}

\section{Introduction}

A theory of planet formation should be able to explain the variety of planetary systems discovered within a coherent framework.
In particular, the presence of gas giants poses a fundamental constraint.Within the core accretion scenario, the interstellar dust grains need to grow from $\mu m$ size to a \unit[$10 - 20$]{M$_\oplus$} planetary core before gas accretion sets in.
This growth must happen before the star can photoevaporate the gas disc,
which occurs on a timescale of \unit[$\sim 3$]{Myr} \citepads{2008PhST..130a4024H}.
Moreover, during the growth period, planets are embedded in the ambient disc and their orbital evolutions are determined by planet-disc and planet-planet interactions.
Some planets may end up accreted onto the star or ejected from the system if no other physical mechanisms intervene to stop them \citepads[see e.g.][]{2012MNRAS.422L..82A,2015MNRAS.450.3008E}.
Planet formation and evolution are determined by the structure of the protoplanetary discs in which they form.
The observations of these discs can give some information about their masses, rotation, and density profile
\citep{2011ARA&A..49...67W}.
The observed diversity in the sample of extrasolar planets indicates that the evolution of a planet may depend on variations in the initial conditions or random (external or internal) events occurring during this crucial phase.
An important initial condition is the stellar environment of the growing planetary system, which can strongly affect the disc lifetime by tidally truncating the outer regions of its birth disc and the dynamical evolution of the planetary system \citepads[see e.g.][]{2015A&A...583A.133P}.

We can place some constraints on the initial conditions and the giant planet formation models by studying their current physical and chemical properties. The Galileo mission measured the abundances of various elements in the outer layers of Jupiter.
\citetads{2003NewAR..47....1Y} found that they were in the range of $2-4$ times solar, with a predicted core mass in the range $\unit[0 - 18]{M_\oplus}$, strongly dependent on the assumed equation of state \citepads{2010SSRv..152..423F}.
The internal composition has also been derived for hot Jupiters such as HAT-P-13~b, where \citetads{2016ApJ...821...26B} used the analysis of secondary eclipses of the planet to infer a core mass of $M_\mathrm{c} < \unit[25]{M_\oplus}$ with a most likely value of $\unit[11]{M_\oplus}$.
These observations can be explained by a bottom-up model of planet formation, such as the core accretion model \citepads{1996Icar..124...62P}, which predicts an enriched solid composition respect to the solar one.
Within this framework, we are interested in studying the process that can explain how the minimum solid core mass, necessary to rapidly build up a massive gaseous envelope, can be accreted within the disc lifetime.

The solid materials accreting onto the forming planetary core can have different origins based on the local size distribution (and Stokes number) of the solid disc.
One solid reservoir consists of the planetesimals that are gravitationally perturbed by the planetary embryo.
If they can cross the mean motion resonances with the planet and enter into its gravitational influence zone, they can end up being accreted onto it.
This model of solid core accretion via planetesimals can explain a certain class of gas giants within few au from the central star, but the timescales needed to form the observed planets at tens or hundreds of au are prohibitive.

One possibility to overcome this limitation is the rapid accretion of pebble-sized particles \citepads{2010A&A...520A..43O,2012A&A...544A..32L}.
In this context pebbles are centimetre-to-meter-sized objects that strongly interact with the gas via drag force.
Planetary embryos with an increased gas density in their proximity have an enhanced sphere of influence to accrete solid particles, and they can interact with an higher flux of pebble-sized particles owing to their fast drift speed.
Adopting this accretion channel, the timescale of giant planet formation can be lowered, making the build up of gas giants at tens of au possible, if a significant reservoir of pebbles is present.
In 2D hydrodynamical simulations with embedded particles, this fast accretion was confirmed \citepads{2012A&A...546A..18M}.
The limiting factor for this accretion process is given by the need of a continuous resupply of material from the outer disc because the drift timescale of this pebble-sized particles is very short.
The formation of a strong pressure gradient created, for example, by the growing planet can filtrate particles, thereby preventing these particles from reaching the planet or inner parts of the disc; this  process sets in at the so-called pebble isolation mass of the core and can explain a class of the observed transition discs.
In this paper, we address these limiting factors by studying the evolution of a variety of particle sizes in a global turbulent disc, deriving their accretion rates onto the planet, and obtaining new estimates on the pebble isolation mass.

We consider the evolution of particles in discs made turbulent by a purely hydrodynamical process, the vertical shear instability (VSI) as described in \citetads{2013MNRAS.435.2610N}.
Recently, \citetads{2016A&A...594A..57S} showed that the dust dynamics in VSI-turbulent discs, following the gas behaviour, has a drift speed directed inwards at the disc midplane and outwards in its upper layers in a similar way as for global MHD simulations \citepads{2011ApJ...735..122F}.
This is exactly opposite to the meridional flow observed for laminar viscous discs.
This phenomenon can have important effects on the planet formation process, resupplying materials to the outer disc regions and explaining the observed chondrule population in the outer regions of the solar system \citepads{2002A&A...384.1107B}.
Moreover, \citetads{2016A&A...594A..57S} found that the strong vertical motions induced by the VSI were able to collect pebble-sized particles in rings of high surface density and low relative velocity, potentially aiding the planet formation process through 
streaming instability \citepads{2005ApJ...620..459Y,2018MNRAS.473..796A}. In this work, we extend our recent analysis of the planet-disc interaction in a laminar and turbulent disc \citepads{2017A&A...604A..28S} by adding dust particles to the simulations and study their accretion dynamics on the planet.
We perform two sets of models.
In the first set, the dust is embedded in VSI turbulent discs (with no explicit viscosity added) and in the second series corresponding viscous discs models are performed.
This allows us to disentangle the effect of turbulence on the planet-dust interaction and the resulting accretion rate of solid particles.
Recently, \citetads{2017ApJ...847...52X} studied the accretion of pebbles on small cores in turbulent discs driven by the magneto-rotational instability (MRI).
We compare our results to their study.

In Sec.~\ref{sec:dustdyn} we describe the various forces acting on dust particles embedded in the protoplanetary disc, and then focussing in Sec.~\ref{sec:plandisc} on the models for planet-disc interaction and solid particle accretion.
In Sec.~\ref{sec:setup} we describe the adopted set-up for the numerical analysis and the main results obtained in Sec.~\ref{sec:results}.
Finally, we discuss the obtained solid accretion rate onto a growing planetary core in Sec.~\ref{sec:solidacc} and draw the main conclusions in Sec.~\ref{sec:concl}.

\section{Dust dynamics}\label{sec:dustdyn}

We consider a thin vertically isothermal gaseous disc with an embedded  protoplanet of mass $M_\mathrm{p}$ orbiting around a Sun-like star.
Additionally, we follow simultaneously the motion of dust particles of various sizes whose motions are determined
by the star, planet, and turbulent gas.
In the VSI turbulent disc models, the particles experience the normal drag forces due gas-particle interaction; see Sect.~\ref{sec:drag}.
On the other hand, in the viscous disc models, the effect of the underlying turbulence is modelled via additional
stochastic kicks on the particles as described in Sect.~\ref{sec:turb}, in addition to the regular drag forces.

\subsection{Equations of motion}\label{sec:eqdust}

A dust particle immersed in the disc is subject to
\textit{(i)} the gravitational force of the central star and the protoplanet,
\textit{(ii)} the drag force due to the varying velocity between the dust orbiting with a Keplerian speed, and the gas,
    which rotates with a slightly sub-Keplerian speed \citepads{1964PNAS...52..565W}, due to the
   radial pressure gradient that partially supports it against the stellar gravity,
\textit{(iii)} gas turbulent motion, which influences the dust dynamics by radially and vertically spreading small particles;
\textit{(iv)} photophoretic gas pressure and radiation pressure, which we do not take into account since we focus mainly
   on a region close to the disc midplane where the efficiency of these processes is expected to be low,
\textit{(v)} growth/fragmentation (depending on their composition/relative speed) due to collisions between grains 
  \citepads[see e.g.][]{2014prpl.conf..339T}. We do not consider this because we study the dynamics
  of particles of varying sizes remaining agnostic about the dust size distribution.

We define a reference frame in spherical coordinates, centred at the location of the star with mass $M_\star = \unit[1]{M_\sun}$ and co-rotating with constant angular velocity $\vec{\Omega}_\mathrm{f}$, following a protoplanet of mass $M_\mathrm{p}$ and fixed position $\vec{r}_\mathrm{p}$. Then, we can describe the equation of motion of a dust particle of mass $m_\mathrm{d}$ as
\beq
    \label{eq:motion}
    m_\mathrm{d}\ddot{\vec{r}}_\mathrm{d}= \vec{F}_\mathrm{grav} +
    \vec{F}_\mathrm{drag} + \vec{F}_\mathrm{turb} + \vec{F}_\mathrm{nonin}\,.\eeq
In this equation, the first term is the gravitational interaction with the star and
planet,
\beq
    \label{eq:Forcegrav}
    \vec{F}_\mathrm{grav} = -\frac{G M_\star m_\mathrm{d}}
    {|\vec{r}_\mathrm{d}|^3}\vec{r}_\mathrm{d} +
    \frac{G M_\mathrm{p} m_\mathrm{d}}
    {|\vec{r}_\mathrm{p} - \vec{r}_\mathrm{d}|^3}
    (\vec{r}_\mathrm{p} - \vec{r}_\mathrm{d}) \,,
\eeq
the second term is the drag force (see Sec.~\ref{sec:drag}), the third term is
the turbulence force (see Sec.~\ref{sec:turb}), and the last is the non-inertial term imparted to the star by the planet,
\beq
    \label{eq:Forcenonin}
    \vec{F}_\mathrm{nonin} =
    - \frac{G M_\mathrm{p} m_\mathrm{d}}
    {|\vec{r}_\mathrm{p}|^3}\vec{r}_\mathrm{p}\,.
\eeq
We ignore the self-gravity of the disc since we model a low mass disc in which a large planetary core has already formed.

\subsection{Drag force}\label{sec:drag}

The drag force acting on a particle depends strongly on the physical condition
of the gas and the shape, size, and velocity of the particle.
We limit ourselves to spherical particles, for which the drag force always acts in the direction opposite to the relative velocity.
The drag regime experienced by a dust particle is described by three non-dimensional parameters as follows:
\begin{enumerate}
  \item The Knudsen number, $K = \lambda/(2 s)$, is the ratio of two characteristic length scales of the system:
  the mean free path of the gas molecules $\lambda$ and the particle size, where $s$ denotes the particle radius.
  \item The Mach number, $M = v_\mathrm{r}/c_\mathrm{s}$, is the ratio of the relative velocity between
   dust and gas, $\vec{v}_\mathrm{r}$, to the gas sound speed $c_\mathrm{s}$.
  \item The Reynolds number is given by
  \beq
    \Rey = \frac{2 v_\mathrm{r}s}{\nu_\mathrm{m}} \,,
  \eeq
  where $\nu_\mathrm{m}$ is the gas molecular viscosity defined as
  \beq
    \nu_\mathrm{m} = \frac{1}{3}{\left(\frac{m_0\bar{v}_\mathrm{th}}{\sigma}
    \right)} \,,
  \eeq
   and $m_0$ and $\bar{v}_\mathrm{th} = \sqrt{\pi/8} c_\mathrm{s}$ are the mass and mean thermal velocity
  of the gas molecules, and $\sigma$ is their collisional cross section.
\end{enumerate}

\subsubsection{Drag law}\label{sec:generalDrag}

We adopt a law that can model the drag force for a broad range of Knudsen numbers, using the approach
implemented by \citetads{2003A&A...399..297W}, who used a quadratic interpolation between the Epstein and Stokes regimes
\beq
    \vec{F}_\mathrm{drag}= \left(\frac{3{K}}{3{K}+1}\right)^2\vec{F}_\mathrm{drag,E} +
    \left(\frac{1}{3{K}+1}\right)^2\vec{F}_\mathrm{drag,S} \,.
\eeq
For large Knudsen numbers, the first term dominates reducing the drag to the Epstein regime \citepads{1965MNRAS.130...63B,1975ApJ...198..583K},\beq
    \vec{F}_\mathrm{drag,E}= -\frac{4}{3}\pi
    \left(1+\frac{9\pi}{128}{M}^2\right)^{1/2}\rho_\mathrm{g}
    (\vec{r}_\mathrm{p})s^2\bar{v}_\mathrm{t}
    \vec{v}_\mathrm{r} \,,
\eeq
where $\rho_\mathrm{g}(\vec{r}_\mathrm{p})$ is the gas density at the particle location.
For small Knudsen numbers, the second term dominates leading to the Stokes regime
\beq
    \vec{F}_\mathrm{drag,S}=-\frac{1}{2}C_\mathrm{D}\pi s^2\rho_\mathrm{g}
    (\vec{r}_\mathrm{p}) v_\mathrm{r} \vec{v}_\mathrm{r} \,,
\eeq
where the drag coefficient $C_\mathrm{D}$ for low Mach numbers is \citepads{1972fpp..conf..211W,1977MNRAS.180...57W}
\beq
    C_\mathrm{D} \simeq
    \begin{cases}
        24\,\Rey^{-1} & \Rey < 1 \\
        24\,\Rey^{-0.6} & 1 < \Rey < 800 \\
        0.44 & \Rey > 800 \,.
    \end{cases}
\eeq
For more information, see \citetads[][and references therein]{2015A&A...584A.110P}.

\subsubsection{Stopping time}\label{sec:Tstop}

A fundamental parameter to determine the strength of the drag force is the
stopping time, $t_\mathrm{s}$, defined as
\beq
    \vec{F}_\mathrm{drag} = -\frac{m_\mathrm{d}}{t_\mathrm{s}}
    \vec{v}_\mathrm{r} \,.
\eeq
It approximates the timescale on which the embedded gas particles approach the velocity of the gas.
In the Epstein regime, the stopping time takes the form
\beq
    \label{eq:stop}
    t_\mathrm{s}=\frac{s\rho_\mathrm{s}}{\rho_\mathrm{g}\bar{v}_\mathrm{th}} \,,
\eeq
where $\rho_\mathrm{s}$ is the internal particle density.
It is also useful to derive a dimensionless stopping time (or, hereafter, Stokes number) as
\begin{equation}
\label{eq:stokes}
    \tau_\mathrm{s}=t_\mathrm{s}\Omega_\mathrm{K}(\vec{r}) \,,
\end{equation}
where $\Omega_\mathrm{K}$ is the Keplerian orbital frequency. The Stokes number $\tau_\mathrm{s}$ (sometimes abbreviated by $\St$) describes the effect of a drag force acting on a particle independent of its location within the disc.
With our definition of the stopping time in eq.~(\ref{eq:stop}) the Stokes number is defined in the midplane of the disc.

\subsection{Turbulence}\label{sec:turb}

Turbulence in the gas acts to stir up well-coupled solid particles, preventing the settling process into a thin layer at the disc midplane.
In general, the source of this turbulence is unknown (either driven by MHD or purely hydrodynamic processes as in our case), but it is responsible for both angular momentum and particle transport within the disc \citepads{2010apf..book.....A}.
By equating the gravitational force in the vertical direction $|\vec{F}_\mathrm{grav}|_z$ with the drag force $|\vec{F}_\mathrm{drag}|_z$,
we can derive a characteristic settling speed
\begin{equation}
    v_\mathrm{set}=t_\mathrm{s}\Omega_\mathrm{K}^2 z \,.
\end{equation}
The condition for which the turbulence strength can counteract the vertical settling of small dust particles is then obtained by comparing the settling time
\begin{equation}\label{eq:set}
    t_\mathrm{set} = \frac{z}{v_\mathrm{set}}
    = \frac{1}{t_\mathrm{s}\Omega_\mathrm{K}^2} \,,
\end{equation}
to the time $t_\mathrm{diff}$ the turbulence needs to erase the spatial gradients in the particle concentration
\begin{equation}\label{eq:diff}
    t_\mathrm{diff} = \frac{z^2}{D_\mathrm{d}}  \,,   
\end{equation}
where $D_\mathrm{d}$ is the turbulent diffusion coefficient of the particles (dust).
If one assumes that $D_\mathrm{d}$ equals, to a first approximation, the diffusion coefficient for the gas
$D_\mathrm{g}$ and that we can write $D_\mathrm{g} \simeq \alpha c_\mathrm{s} h$, assuming that the turbulence acts like
an effective viscosity \citepads{1973A&A....24..337S}, then one can
derive the minimum $\alpha$ value required to prevent dust settling at one scale height, $z=h$
\begin{equation}
    \alpha \ga \tau_\mathrm{s}\,.
\end{equation}
In the following we use a more complex turbulent diffusion model that distinguishes between
$D_\mathrm{d}$ and $D_\mathrm{g}$.

\subsubsection{Turbulent diffusion model}\label{sec:TDM}

The source of turbulence in planet-forming discs is unknown. It can depend strongly on the environment, and different sources might be dominant in the various regions and during the evolution of the disc. In the laminar disc simulation, we do not consider the origin of the turbulence and use a simplified turbulence diffusion model to evolve the dust population. The basic idea is to mimic turbulent transport as a diffusive process (through a Brownian motion) \citepads{Dubrulle1995Icar..114..237D,2011ApJ...737...33C,2007Icar..192..588Y} with a stochastic term in the equation of dust motion to account for the kicks induced by the turbulent gas velocity field. We model the kick on the particle position as a random Gaussian variable $\delta r_\mathrm{d,T}$ with mean $\langle\delta r_\mathrm{d,T}\rangle$ and variance $\sigma_\mathrm{d,r}^2$ depending on the dust diffusion coefficient $D_\mathrm{d}$ as follows:
\begin{equation}
    \delta r_\mathrm{d,T} =\left\{
    \begin{array}{l}
        \langle\delta r_\mathrm{d,T}\rangle = \frac{D_\mathrm{d}}{\rho_\mathrm{g}}\frac{\partial \rho_\mathrm{g}}{\partial x} dt\\[2mm]
        \sigma_\mathrm{d,r}^2 = 2 D_\mathrm{d} dt\,,
    \end{array}\right.
\end{equation}
where $dt$ is the time step and $\partial/\partial x$ is the spatial derivative along the considered direction.
The relation between particle and gas diffusion can be written as
\begin{equation}
    D_\mathrm{d} = \frac{D_\mathrm{g}}{\Sch}\,,
\end{equation}
where $\Sch$ is the Schmidt number \citepads{2007Icar..192..588Y}
\begin{equation}
  \Sch = \frac{1+\Omega_\mathrm{K}^2 \tau_\mathrm{s}^2}{1+4\Omega_\mathrm{K}\tau_\mathrm{s}}\,.
\end{equation}
This prescription for the turbulent diffusion assumes that the diffusion coefficients in the vertical and radial direction
are identical. This is a crude assumption as we have shown in \citetads{2017A&A...599L...6S} because the $\alpha$
parameter can differ in the radial and vertical direction by more than two orders of magnitude;
see also the discussion in \citetads{2007Icar..192..588Y}.
Nevertheless, we use this simplified description to model the vertical and radial spread of dust particles in
the viscous disc simulations and find
results comparable to those created self-consistently by VSI turbulence as we show later.

\section{Planet-solid disc interaction}\label{sec:plandisc}

\subsection{Particle accretion}\label{sec:accpart}

The idea that gas plays a pivotal role in the accretion of solids by planetary cores was first introduced in the Kyoto model \citepads{1977PASJ...29..163H,1981PThPS..70...11N,1983Icar...54..361N}. The basic concept was that the orbital decay experienced by planetesimals is size dependent, occurring at a lower rate for larger bodies. In this way, a large embryo can grow as drag feeds it with dust and small planetesimals.
Later, \citetads{1985Icar...62...16W} found that orbital resonances with the growing core can effectively filter a significant fraction of planetesimals for which the drag force is not strong enough to allow them to cross those stable regions. However, since eccentricities are pumped up at resonances, collisions between large planetesimals become more frequent, increasing the fraction of smaller bodies that can cross the resonances and accrete onto the planetary core.

On the other hand, \citetads{1993Icar..106..288K} showed that even if a body is small enough to cross all the resonances, it can avoid being accreted. The impact probability typically ranges between $10 - 40\%$ but can be higher if the core possesses an extended atmosphere \citepads{2014Icar..241..298D}. More generally, the accretion rate is inversely proportional to the strength of the drag force and the inclination of the planetesimal. Moreover, \citetads{1993Icar..106..288K} found that for cores with mass ratio $q=M_\mathrm{p}/M_\star > 10^{-5}$, the material approaching the planet can be captured into a stable orbit around the planet, thereby forming an accretion disc around it.

This strong perturbation in the local environment of the protoplanet creates pressure gradients that impact the evolution of dust and planetesimals \citepads{2004A&A...425L...9P,2007A&A...462..355P}.
In particular, small protoplanets, depending on their surface and temperature profiles, can carve a gap in the dust disc even
if there is no gas gap \citepads{2015A&A...584A.110P,2016MNRAS.459.2790R,2016MNRAS.459L...1D}.

\subsection{Resonances}\label{sec:res}

A particle that migrates within the disc feels a stronger (regular) gravitational interaction with a planetary companion when it reaches specific locations in the disc where its mean motion $n_\mathrm{d}=2\pi/T$, where $T$ is its orbital period, is a multiple of the planet mean motion $n_\mathrm{P}$
\begin{equation}
  \frac{n_\mathrm{d}}{n_\mathrm{P}} = \frac{(l+m)}{l}\,,
\end{equation}
where $l$ and $m$ are integer numbers. These are called mean motion orbital resonances (MMR), where $m$ gives the order of the resonance. These MMRs can effectively excite the eccentricity and inclination of particles, potentially halting their drift process. Their strength grows for decreasing values of $m$ and increasing $l$. Thus, focussing on first order MMR ($m=1$), the larger $l$, the smaller the particles that can be stopped from accreting onto the planet. The resonances are yet not able to halt all the particles because they become more and more closely spaced as $l$ grows until the point at which they overlap leading to a chaotic behaviour of the dust particles that can cross the higher order resonances \citepads{1980AJ.....85.1122W}. The minimum size $s_\mathrm{min}$ of a particle for which the resonant perturbations due to a planet with mass ratio $q$ are stronger than the drag force is \citepads{1985Icar...62...16W}
\begin{equation}
    s_\mathrm{min}=\frac{\rho_\mathrm{g} h a_\mathrm{d}}{3
    \rho_\mathrm{d} q C(l) l^{3/2}}\,,
\end{equation}
where $C(l)$ is an increasing function of $l$, $a_\mathrm{d}$ is the particle semi-major axis, and the region of chaotic behaviour close to the planet location starts at \citepads{1989Icar...82..402D}
\begin{equation}\label{eq:rminres}
    |r-a_\mathrm{d}| \simeq 1.5 q^{2/7}\,.
\end{equation}
This relation depends strongly on the local gas properties. When the planet opens up a gap in the gaseous disc, reducing the gas surface density, the particle Stokes number increases; thus the inner resonances can halt a larger fraction of incoming particles, which are less coupled to the gas.

For a planet with $q=10^{-4}$ \citetads{2004A&A...425L...9P} found three visible regimes.
Particles with Stokes number less than $\tau_\mathrm{s} \simeq 0.1$ are well coupled to the gas, and they always reach the planet surface.
On the other hand, particles with $\tau_\mathrm{s} > 10$ are trapped in external resonances, and their accretion rate is very low.
Finally, the intermediate regime is reaching the co-orbital region of the planet, but not all of them are accreting as predicted by \citetads{1993Icar..106..288K}.

\section{Set-up}\label{sec:setup}

We used the \textsc{pluto} code \citepads{2007ApJS..170..228M} and modified it to take into account the evolution of partially coupled particles.
The main parameters of the reference simulations are summarised in
Tab.~\ref{Tab:sum}.
The simulations analysed in this work are the same as in \citetads{2017A&A...604A..28S}, where we analysed the dynamics of a planet embedded in a VSI turbulent disc without particles, so we briefly describe the set-up, focussing only on the dust part.
For a more detailed description of the initial conditions of the gaseous disc, see \citetads{2017A&A...604A..28S}.

\subsection{Gas component}\label{par:gasdisc}

The initial disc profile is axisymmetric and extends from 2.08~au to 13~au ($0.4$ to $2.5$ in code units,
where the unit of length is \unit[$5.2$]{au}).
The gas moves with azimuthal velocity given by the Keplerian speed around a
$\unit[1]{M_\sun}$ star, corrected for the pressure term and rotational velocity of the coordinate system that rotates here with the orbital speed of the planet.
The total disc mass is $\unit[0.01]{M_\sun}$ and the density distribution, created by force equilibrium, is given in cylindrical coordinates ($R,Z,\phi$) by
\begin{equation}
    \rho_\mathrm{g}(R,Z)=\rho_{\mathrm{g},0}\, \left(\frac{R}{R_\mathrm{p}}\right)^{p}\, \exp{\left[\frac{G M_\mathrm{s}}{c_\mathrm{s}^2}\left(\frac{1}{r}-\frac{1}{R}\right)\right]}\,,
    \label{eq:surfprof}
\end{equation}
where $\rho_{\mathrm{g},0}$ is the gas midplane density at $R = 1$ and $p=-1.5$ is the density exponent.
In our case $\rho_{\mathrm{g},0} = \unit[2.07 \cdot 10^{-11}]{\mathrm{g/cm}^{3}}$ such that the vertically integrated surface density  at $R = 1$ is $\Sigma_{\mathrm{g}} = \unit[200]{\mathrm{g/cm}^{2}}$.
The disc is modelled with a locally isothermal equation of state, and we assume a constant aspect ratio $H/R = 0.05$, which corresponds to a radial temperature profile with an exponent $q=-1$ and $T(R_\mathrm{p}) = \unit[121]{K}$.
For the inner and outer radial boundary, we apply reflective conditions, while outflow conditions are implemented for the vertical boundaries and periodic conditions in the azimuthal direction.
We perform two sets of simulations.\ One set has an inviscid disc in which the source of turbulence is given entirely by the VSI and the other uses a viscous disc in which the viscosity is given by
$\nu = 2/3 \alpha c_\mathrm{s} H,$ where we use a constant $\alpha$ viscosity as derived from the VSI simulation, which is $\alpha = 5\cdot10^{-4}$
\citepads{2017A&A...604A..28S}.

\subsection{Dust component}\label{par:dustdisc}

The solid fraction of the disc is modelled with $10^6$ Lagrangian
particles divided into ten size bins as reported in Table~\ref{Tab:sum}.This approach has the great advantage of modelling a broad range of Stokes numbers (see eq.~\ref{eq:stokes}) self-consistently using the same model particles.
The trade-off is that in the regions of low density, the resolution of the dust population is lower. However, for our study this is not a problem since we are mainly interested in the dynamical evolution of dust particles; thus we do not take into account collisions between particles or the backreaction of the dust onto the gas.
We study particles with sizes from $\unit[0.1]{mm}$ up to
$\unit[1]{km}$ and internal density $\rho_\mathrm{d}=\unit[1]{g {cm}^{-3}}$.
The particle sizes and corresponding Stokes numbers are quoted in Table~\ref{Tab:sum}, where the Stokes numbers are evaluated at the planet location.
The particle sizes are chosen to cover a wide range of different dynamical behaviour.
The initial surface density profile of the dust particles is
\begin{equation}
    \Sigma_\mathrm{d}(r) \propto  R^{-1}\,.
\end{equation}
This choice was made to have a larger reservoir of particles in the outer disc.
This particle distribution leads to equal number particles in each radial ring as the grid is spaced logarithmically in the radial direction.
The dust particles are placed initially at the disc midplane in the disc model with active VSI driven turbulence because the particle stirring is obtained via the turbulent mechanism. For the laminar disc, we start with a vertical distribution given by the local disc scale height and the dust diffusion coefficient. By comparing eq.~(\ref{eq:set}) and eq.~(\ref{eq:diff}) we find that, for our initial profiles, particles larger than $\unit[1]{mm}$ are going to settle to the disc midplane.
The particles are introduced at the beginning of the simulation, and they are evolved with two different integrators depending on their Stokes numbers.
Following the approach by \citetads{2014ApJ...785..122Z}, we adopt a semi-implicit leapfrog-like (drift-kick-drift) integrator in spherical coordinates for larger particles and a fully implicit integrator for particles well coupled to the gas. We  include in Appendix~\ref{sec::integrators} the detailed implementation of the two integrators. We do not consider the effect of the disc self-gravity on the particle evolution.
Particles that leave the computational domain at the inner boundary are re-entered at the outer boundary.
Accreted particles are flagged but are otherwise kept in the simulations.
\subsection{Planets}\label{par:planets}

We embed a planet, with a mass in the range $\unit[[5, 10, 30, 100]]{M_\oplus}$, orbiting a solar mass star on a circular orbit with semi-major axes $a_\mathrm{p}=1$ in code units (\unit[$5.2$]{au}).
The planet does not migrate and its mass is kept fixed. To prevent a singularity close to the planet location, its gravitational potential is smoothed with a cubic expansion inside a sphere centred on the planet location with a radius given by the smoothing length $d_\mathrm{rsm}=0.5 R_\mathrm{H}$ (\citetads{2006A&A...445..747K}, \citetads{2017A&A...604A..28S}), where $R_\mathrm{H}$ denotes the radius of the Hill sphere
\beq
\label{eq:R_Hill}
       R_\mathrm{H} = R_\mathrm{p} \, \left(\frac{1}{3} q \right)^{1/3} \,.
\eeq
After the dust component has been evolved for $\unit[20]{orbits}$ in the computational domain, the planetary mass is slowly increased over additional $\unit[20]{orbits}$ to allow for a smooth initial phase. Each simulation was run over $200$ orbital periods of the planet when the disc structure had reached a quasi-stationary state.

\begin{table}[tb]
    \caption{Model parameter}
    \label{Tab:sum}
    \centering
    \begin{tabular}{l | c}
        \hline\hline
        Parameter & model \\
        \hline
        Radial range [$5.2\ \mbox{au}$] & $0.4$ - $2.5$  \\
        Vertical range [H, $\theta$] & $\pm 5$, \, $76^\circ -104^\circ$ \\
        Phi range [rad] & $0$ - $2\ \pi$   \\
        Radial grid size & $600$  \\
        Theta grid size & $128$  \\
        Phi grid size & $1024$   \\
        Planet masses [M$_\oplus$] & $5$, $10$, $30$, $100$   \\
        \hline
        Particle sizes [cm] & $0.01$, $0.1$, $1$, $10$, $30$, \\
        ($10^5$ in each bin) & $100$, $300$, $10^3$, $10^4$, $10^5$  \\
      Corresponding  & $7.79 \cdot 10^{-5}$, $7.79 \cdot 10^{-4}$, 0.0078 \\
      Stokes number &  0.082, 0.27, 1.23,  6.91,  67.2 \\
      for the 10 bins   &   377, 7670 \\
        \hline
    \end{tabular}
\end{table}

  \begin{figure}[htbp]
      \centering
      \includegraphics[width=\columnwidth]{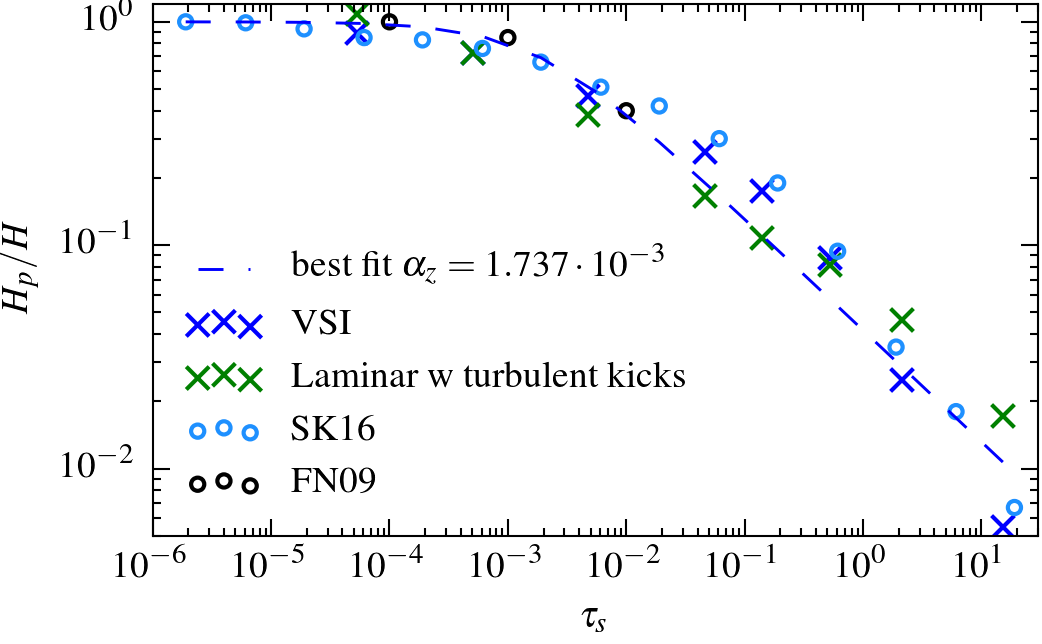}
          \caption{Measured particle scale height, $H_\mathrm{p}$, in units of the gas scale height, $H$, as a function of the particle Stokes number for the simulations of the turbulent VSI, and the laminar disc plus stochasistic kicks at $R=1.8 r_\mathrm{p}$, averaged between $150$ and $200$ orbital periods.
          Shown are the results of the runs used in this paper (labelled with blue and green crosses), and of \citetads{2016A&A...594A..57S} and \citetads{2009A&A...496..597F} indicated with light blue and black circles, respectively.
          Additionally, we overplot the fit of the VSI particle scale heights by \citetads{2007Icar..192..588Y} (see equation~\ref{eq:dustscale}).
      }
      \label{fig:scaleheight}
  \end{figure}

\begin{figure*}[htbp]
    \centering
    \includegraphics{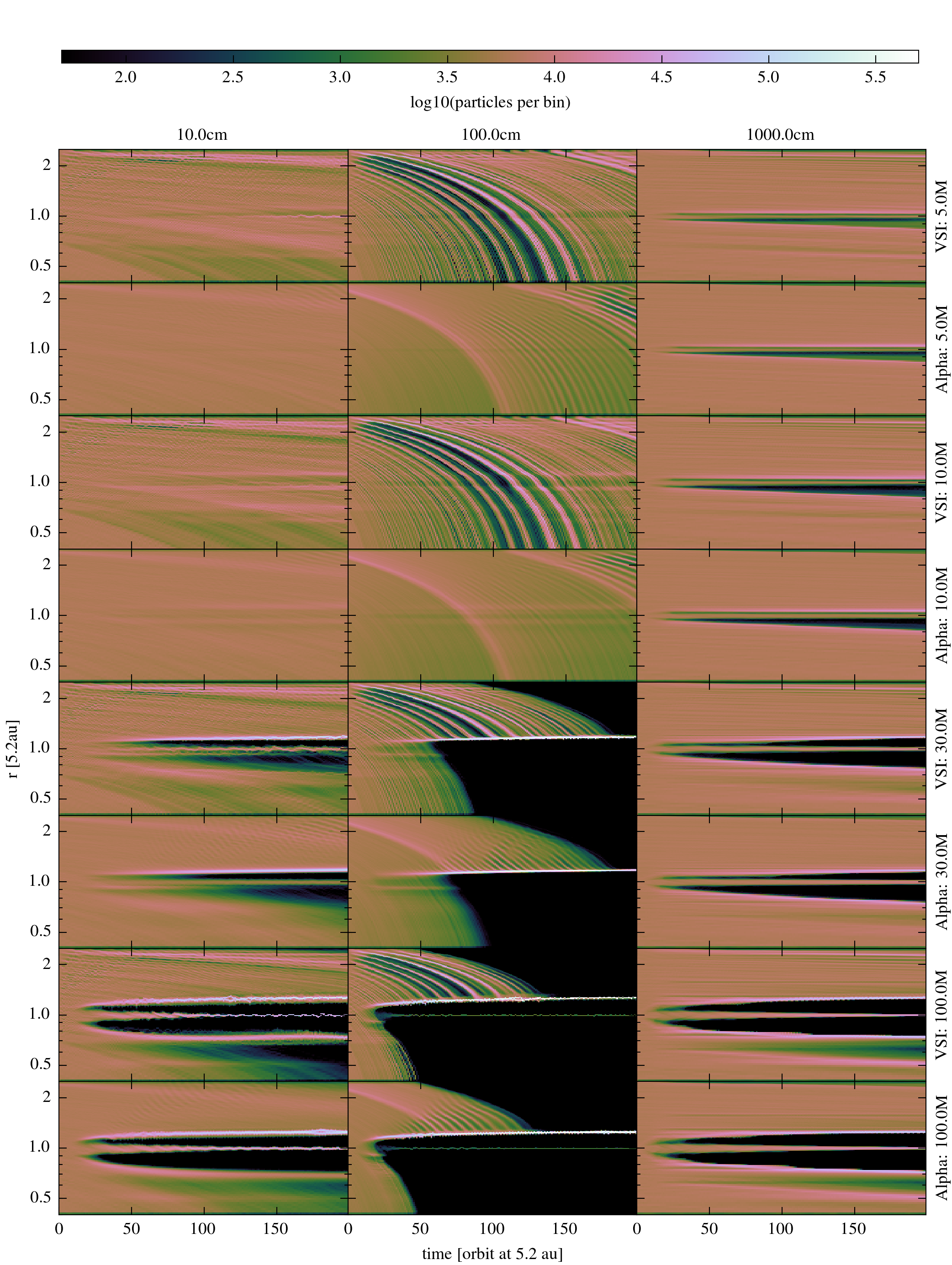}
    \caption{Spatial distribution of the dust particles as a function of time for the different planetary masses and for three representative particle sizes. The Stokes numbers from left to right are $0.08$, $1.23,$ and $67.2$.}
    \label{fig:histTime}
\end{figure*}

\begin{figure*}[htbp]
    \centering
    \includegraphics{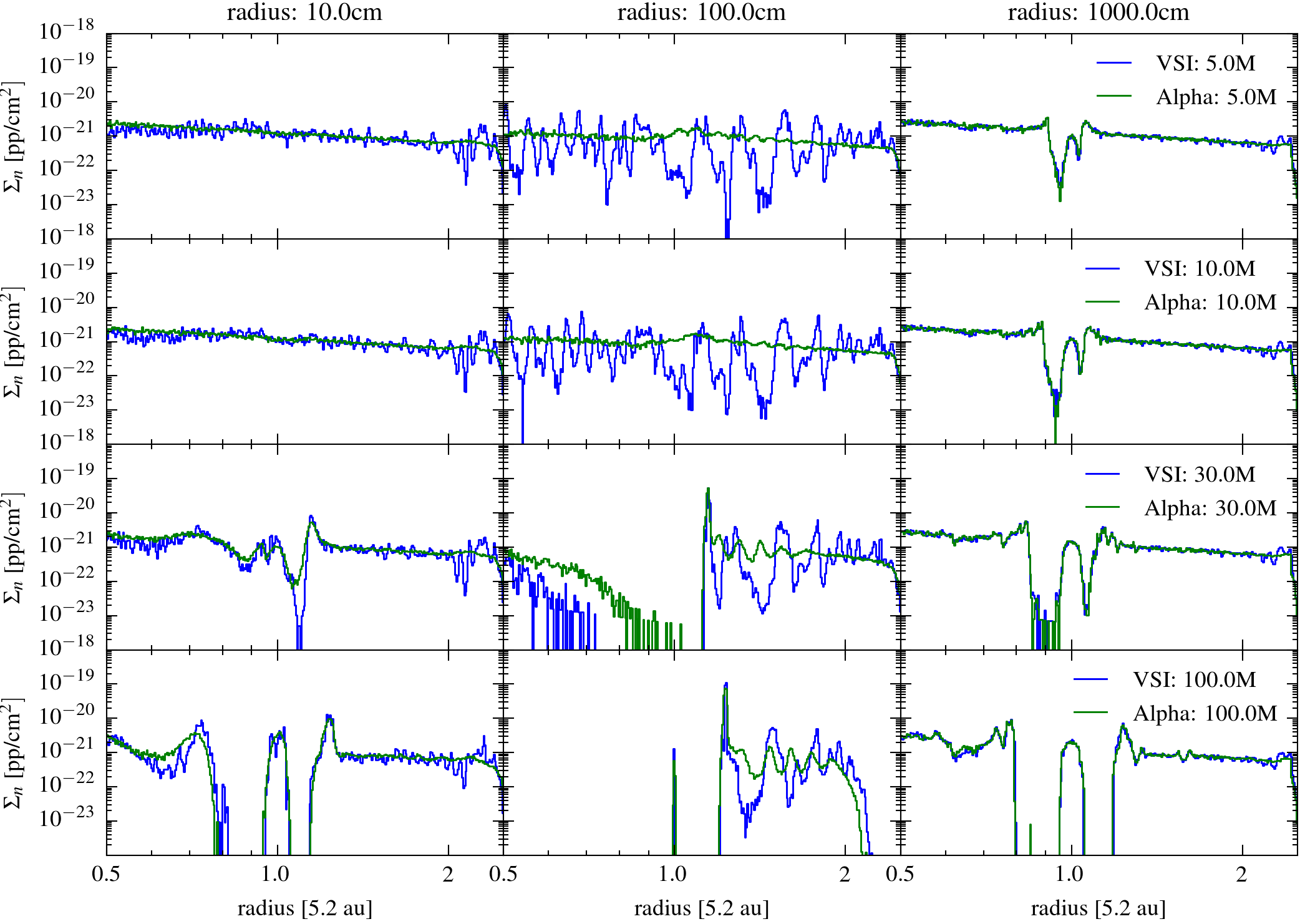}
    \caption{Histogram of the dust surface number density distribution as a function of radius after $200$ planetary orbits for three representative dust sizes in the turbulent (blue) and laminar (green) case.
    The Stokes numbers from left to right are $0.08$, $1.23,$ and $67.2$.
   }
    \label{fig:histRadial}
\end{figure*}

\section{Global dust motion}\label{sec:results}
In this section, we analyse the overall behaviour of the dust particles in the presence of the planet in combination with the disc turbulence.
Of particular interest are the changes in the spatial distribution of the particles as induced by the planet.
However, before we focus on the action of the planet we comment briefly on the dust dynamics in the disc.

\subsection{Vertical dust distribution}
\label{subsec:dust_distribution}
The action of the turbulence in the disc works against the tendency of the dust particles to settle towards the midplane of the disc and leads to a vertical spreading of the particles.  
Concerning this particle stirring in turbulent discs we present in Fig.~\ref{fig:scaleheight}
the vertical scale height of the particles (in units of the gas scale height $H$) as a function of their Stokes number, $\tau_\mathrm{s}$ for various models.
The two cases studied in this work are shown by the blue and green crosses for the VSI turbulent case and the laminar disc with
stochastic particle kicks, respectively.
We analysed the particle distribution for the $5 M_\oplus$ case at a radius of $1.8 r_\mathrm{p}$ averaged between $150$ and $200$ orbital periods.
From our previous work in \citetads{2016A&A...594A..57S} we know that the timescale for spreading the particles vertically in the presence of
fully developed VSI turbulence is about $100$ orbital periods, so near the end of our simulations ($150-200$ orbits)
the particle distribution has reached a quasi-stationary state.
Furthermore, we checked the particle vertical distribution from $100$ and $150$ orbital periods at the same distance
and found the same profile, confirming that an equilibrium was reached.
Additionally, we show the results of \citetads{2016A&A...594A..57S} for locally isothermal discs using the data taken from their Table~1
(labelled SK16) and \citetads{2009A&A...496..597F} who studied particle settling in global ideal MHD disc displaying MRI turbulence (labelled FN09).

Overplotted to the data is an approximation by \citetads[][their equation (28)]{2007Icar..192..588Y}, which can be written as
(neglecting a correction factor of order unity) 
\beq
   \frac{H_\mathrm{p}}{H} = \sqrt{ \frac{\alpha_z}{\alpha_z + \tau_\mathrm{s}} } \,,
\label{eq:dustscale}
\eeq
where $\alpha_z$ measures the vertical diffusion of the gas; see \citetads{2007Icar..192..588Y}.
In Fig.~\ref{fig:scaleheight} we use $\alpha_z = 1.737\cdot10^{-3}$ for the fit. 
Equation (\ref{eq:dustscale}) accounts for the fact that for small $\tau_\mathrm{s}$ the particles are well coupled
to the gas and the two scale heights agree, $H_\mathrm{p} = H$, while
larger particle settle more to the midplane of the disc and have a smaller thickness. For large $\tau_\mathrm{s}$ the
slope becomes $\propto \tau_\mathrm{s}^{-0.5}$ as can be inferred from eq.~(\ref{eq:dustscale}). 

For the small particle sizes our distribution is similar to that of 
\citetads{2009A&A...496..597F}.
For their
investigated particle sizes with $\tau_\mathrm{s}$ = ($10^{-4}, 10^{-3}, 10^{-2}$), they find a scaling 
$H_\mathrm{p}/H \propto \tau_\mathrm{s}^{-0.2}$ in rough agreement with our findings. 
For the larger particles
the slope becomes steeper than the expected $\propto \tau_\mathrm{s}^{-0.5}$ scaling
because we have reached the resolution limit in our simulations such that the particle scale height cannot be resolved anymore.

In our previous simulations of particles embedded in VSI turbulent discs \citep{2016A&A...594A..57S}
we find for the mean vertical velocity at 5\,au  
$< v_z^2> = 5 \cdot 10^{-6} v^2_\mathrm{K,1au}$ (normalising to the Keplerian velocity at \unit[1]{au}).
Using this value and $H/r= 0.05$ we find for the mean vertical Mach number $M_z \approx 0.1$.
In \citep{2016A&A...594A..57S} we quote for the (dimensionless) eddy turnover timescale
$\tau_\mathrm{e} \approx 0.2$. From these we can calculate a vertical diffusion coefficient of
\citep{2007Icar..192..588Y}
\beq
  \alpha_z  = \tau_\mathrm{e} \,{M_z^2} \,.
\eeq
Hence, from \citet{2016A&A...594A..57S}
we find $\tau_\mathrm{e} \approx 0.2,$ which is consistent with the value obtained by the fit for Fig.~\ref{fig:scaleheight}.

\begin{figure*}[htbp]
    \centering
    \includegraphics{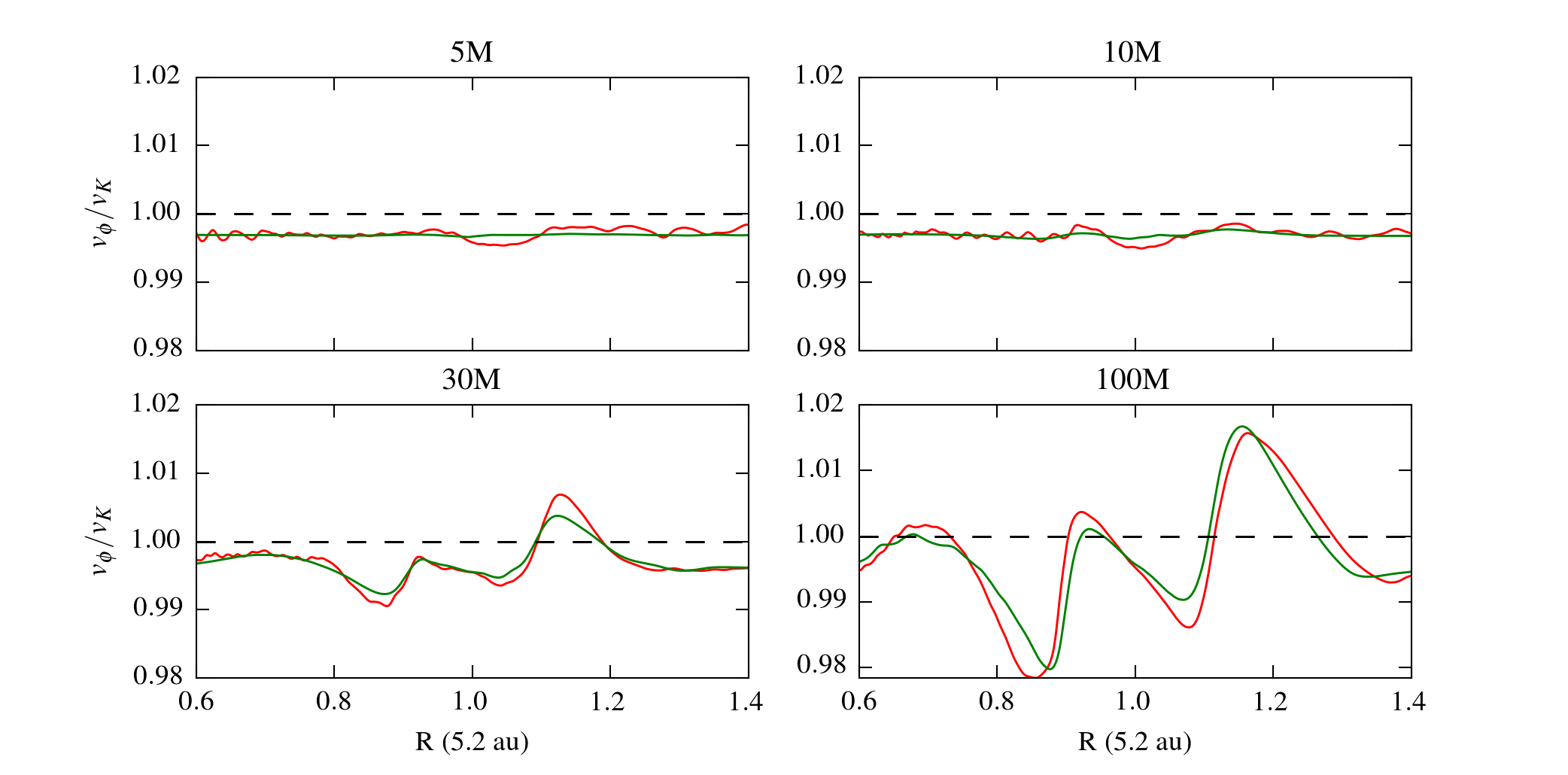}
    \caption{Azimuthal gas speed in units of the Keplerian speed as a function of radius for the different planetary masses and models (VSI = red, alpha-disc = green). When the gas speed becomes super-Keplerian outside the planet location, the dust-filtration process occurs and the pebble isolation mass is reached.}
    \label{fig:vphiKep}
\end{figure*}

\begin{figure*}[htbp]
    \centering
    \includegraphics{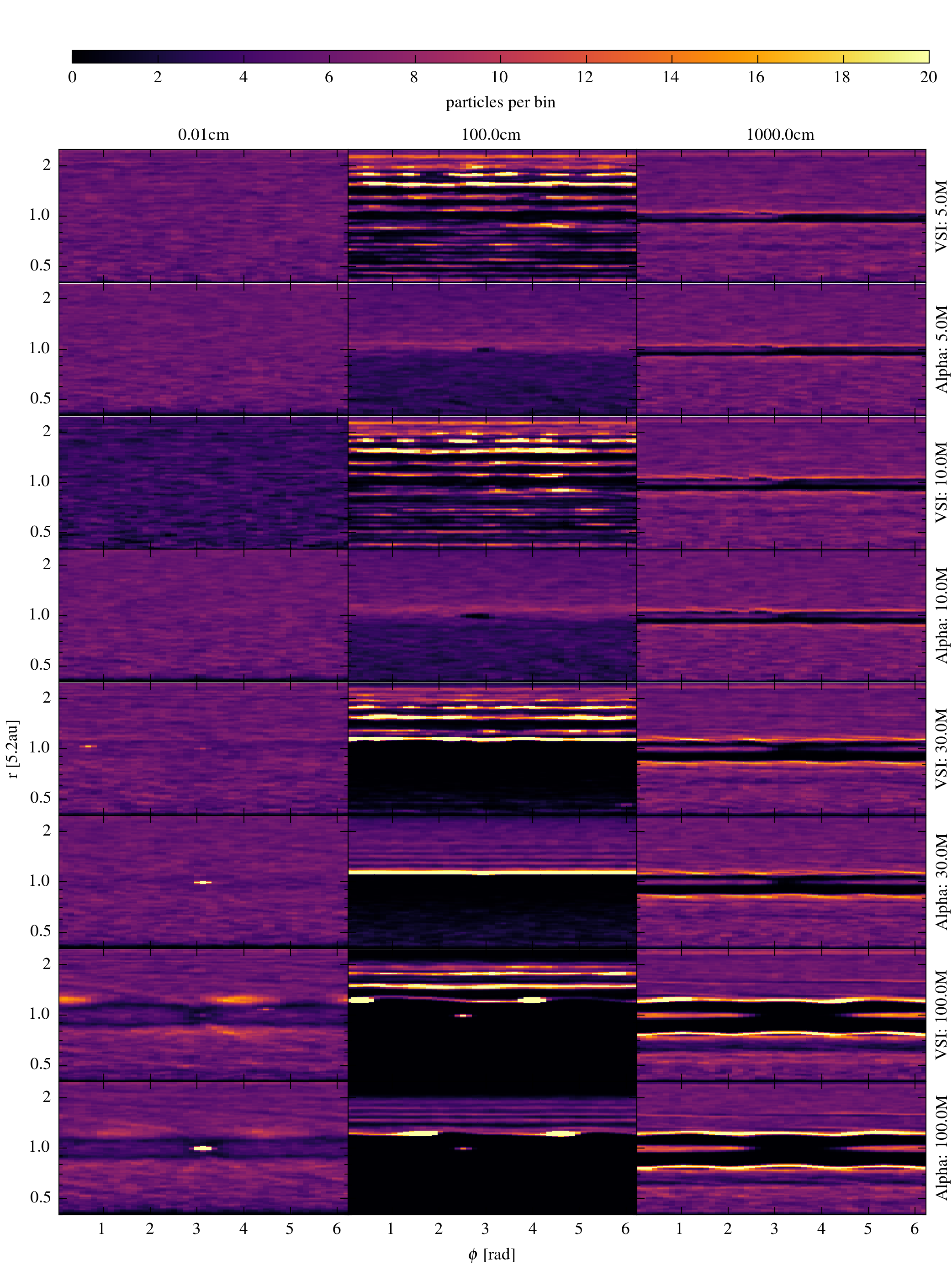}
    \caption{Surface density distribution of the dust particles after \unit[$80$]{planetary orbits} for the different planetary masses and for three representative particle sizes. The Stokes numbers from left to right are $7.79 \cdot 10^{-5}$, 1.23, and 67.2.}
    \label{fig:histTime2}
\end{figure*}

\subsection{Dust filtration}
\label{subsec:filtration}

In Fig.~\ref{fig:histTime} we plot the radial distribution of three representative size particles as a function of time.
Shown are the results for both the viscous $\alpha$- and turbulent VSI-disc models for all planet masses from \unit[$5$]{M$_\oplus$} (top)
to \unit[$100$]{M$_\oplus$} (bottom). The particle sizes increase from left to right from 10.0~cm to 10~m, which corresponds to
the Stokes numbers $8 \cdot 10^{-2}, 1.23$ and $67$, respectively. Clearly visible are the different radial drift velocities of the
particles that are a function of the Stokes number, $\tau_\mathrm{s}$. Indeed, the speeds found in our simulations are in good
agreement with the theoretical expectation of \citetads{Nakagawa1986}, which is given by
\beq
    v_\mathrm{drift} = \frac{\partial \ln p}{\partial \ln R} \, \left(\frac{H}{R}\right)^2
       \frac{u_\mathrm{K}}{\tau_\mathrm{s} +\tau_\mathrm{s}^{-1}} \,
      \equiv \,  - 2 \, \eta \, \frac{v_\mathrm{K}}{\tau_\mathrm{s} +\tau_\mathrm{s}^{-1}} \,.
    \label{eq:radialdrift}
\eeq
Equation~(\ref{eq:radialdrift}) indicates that the maximum speed, reached for $\tau_\mathrm{s} = 1$, is given by $v_\mathrm{drift} = - \eta v_\mathrm{K}$, where
$\eta$ is typically of order $(H/R)^2$ and $v_\mathrm{K}$ is the Keplerian azimuthal velocity.
The results on the drift speed for the VSI turbulent and viscous disc are very similar for all cases studied.

For the low mass \unit[$5$]{M$_\oplus$} planet and small particles one notices small disturbances near the planet
(first two rows on the left), but the planets are not able to stop the particles from crossing their location.
The same behaviour is also found in the \unit[$10$]{M$_\oplus$} case displayed in the second two rows of Fig.~\ref{fig:histTime}.
Focussing on the middle column, we can see the evolution of \unit[$1$]{m} particles, which have a Stokes number of order unity
($\tau_\mathrm{s} = 1.23$). Their drift speed is so high that they can cross the whole computational domain in \unit[$\sim 150$]{orbits},
in agreement with Eq.~(\ref{eq:radialdrift}); see also \citetads{2017A&A...604A..28S}.
A change in the drift speed is also visible as the particles cross the planet co-orbital region because the Stokes number
suddenly increases because of the drop in the gas density.
The only exception is given by the planetesimal-sized objects (\unit[$10$]{m}, $\tau_\mathrm{s} = 67$, right column), which do not feel a strong gas drag,
such that the planetary core can effectively perturb their orbits depleting its co-orbital region.
This regime is described in \citetads{2017MNRAS.469.1932D}, who found that for a Stokes number number greater than a critical value
\beq
    \tau_\mathrm{s,crit} \simeq 2.76 \left(\frac{-\zeta}{1+\epsilon}\right) \simeq 6.83\,,
\eeq
where $\epsilon = 0.01$ is the dust-to-gas ratio, and $\zeta=\partial{\ln{p}}/\partial{\ln{R}} = -2.5$,
the minimum mass to open a gap in the solid disc is
\beq
    M_\mathrm{crit} \simeq 1.38 \left(\frac{-\zeta}{1+\epsilon}\right)^{3/2} \tau_\mathrm{s}^{-3/2} \left(\frac{H}{R}\right)^3\, M_\odot \,.
\eeq
For our parameter space, we find that this transition happens between the particles with $s = \unit[300]{cm}$ ($\tau_\mathrm{s}=6.91$) and a critical mass of \unit[$12$]{M$_\oplus$} and the particles with $s = \unit[1000]{cm}$ ($\tau_\mathrm{s} = 67.2$) and a critical mass of \unit[$0.4$]{M$_\oplus$}.
We see that only the second sample of particles is depleted from the co-orbital region of the small mass planets, confirming their analytical prescription.

The third two rows of Fig.~\ref{fig:histTime} show the particle evolution for a \unit[$30$]{M$_\oplus$} planet. This planetary mass can change the final particle distribution dramatically. A gap is already visible for the \unit[$10$]{cm} particles, while for the particles with Stokes number of order unity (central column) the planet acts as a barrier and can filtrate the dust in the outer disc.
This effect is due to the formation of a pressure maximum beyond the planet where small particles are trapped \citepads{2006A&A...453.1129P}.
After \unit[$\sim 100$]{orbits} the particles are located either in the pressure bump close to the planet position or in the outer disc.
This dust filtration leads to a strong reduction of particles inside the planetary orbit, and hence the number of particles leaving through the inner radius is strongly diminished.
As those particles are re-entered at the outer boundary, eventually this results in a shut-off of the flow of particles in the outer disc.
We decided not to have a constant inflow of particles because they would only end at the pressure bump as all the others.
The particle concentration could become a sweet spot to have a second generation of planets due to streaming instability, but it is beyond the purpose of this paper to study this high dust density regions in more detail.

For the planetesimal-sized objects, the \unit[$30$]{M$_\oplus$} planet can open a deeper and wider gap compared to the small mass cases.
Only planets greater than \unit[$10$]{M$_\oplus$} are able to filter the pebble-sized particles efficiently. This result confirms the value obtained by \citetads{2014A&A...572A..35L} in which they defined the pebble isolation mass around \unit[$20$]{M$_\oplus$} for similar initial disc condition.
As we have seen, a planet that modifies the pressure profile in the disc can effectively stop the inward drift of certain size of dust particles.
In a related scenario, such a dust filtration is believed to explain the observation of a class of transition discs (Type 2), where the dust is highly depleted in the inner region of the protoplanetary disc while the gas accretion rate onto the star remains high \citepads[for a recent review on the subject see][]{2017arXiv170400214E}.
The last two rows show the particle evolution for the \unit[$100$]{M$_\oplus$} planet where the dynamical behaviour of the dust particles is very similar to the \unit[$30$]{M$_\oplus$} case but the gap opens earlier, and so the simulations reach a stationary state on a shorter timescale.
%
\begin{figure*}[htbp]
    \centering
    \includegraphics{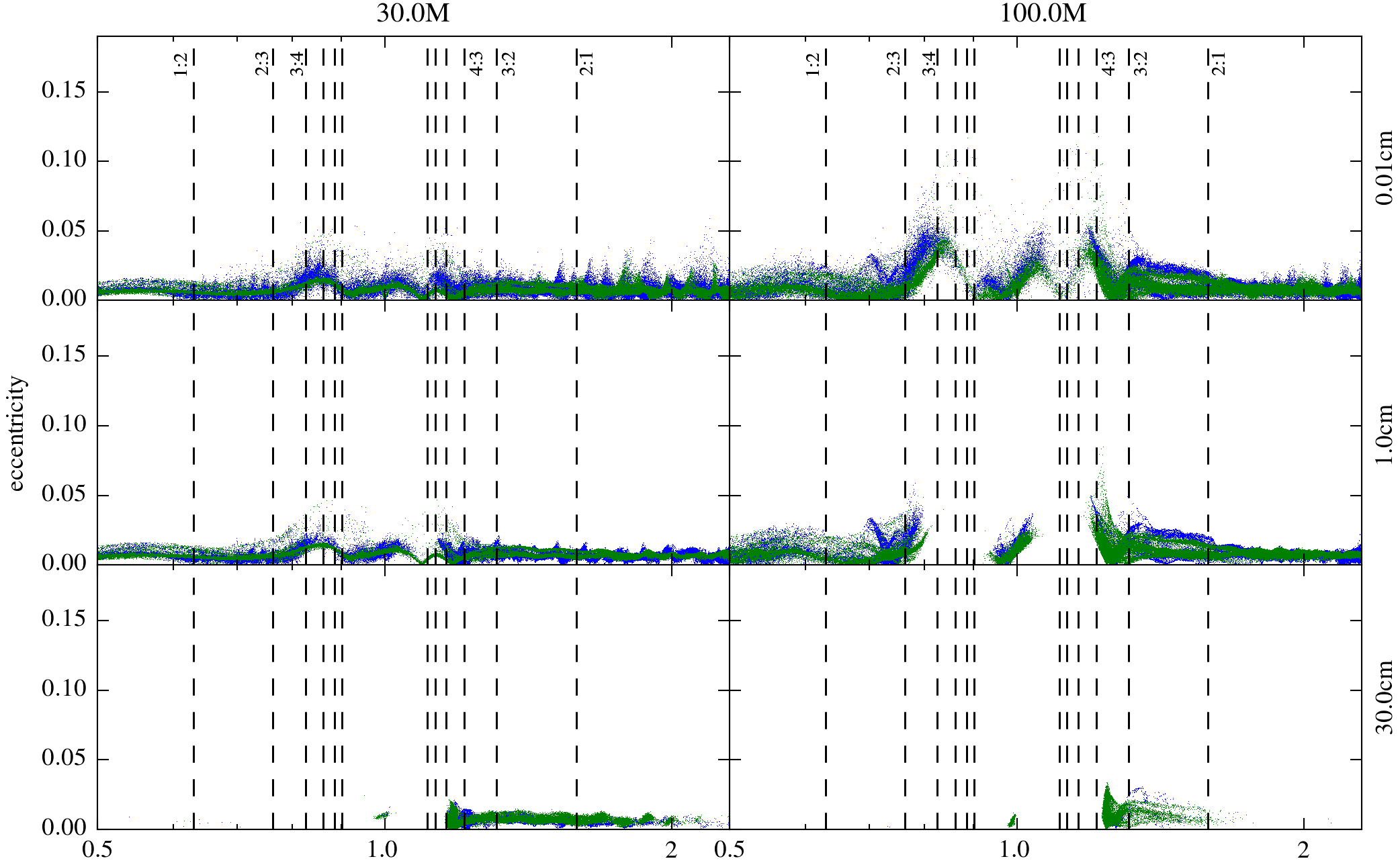}
        \caption{Eccentricity distribution for $3$ representative size particles in the VSI (blue) and alpha-disc (green) model at the end of the simulation. The first six inner and outer first-order MMRs are overlaid.}
    \label{fig:ecc}
\end{figure*}
\begin{figure*}[htbp]
    \centering
    \includegraphics{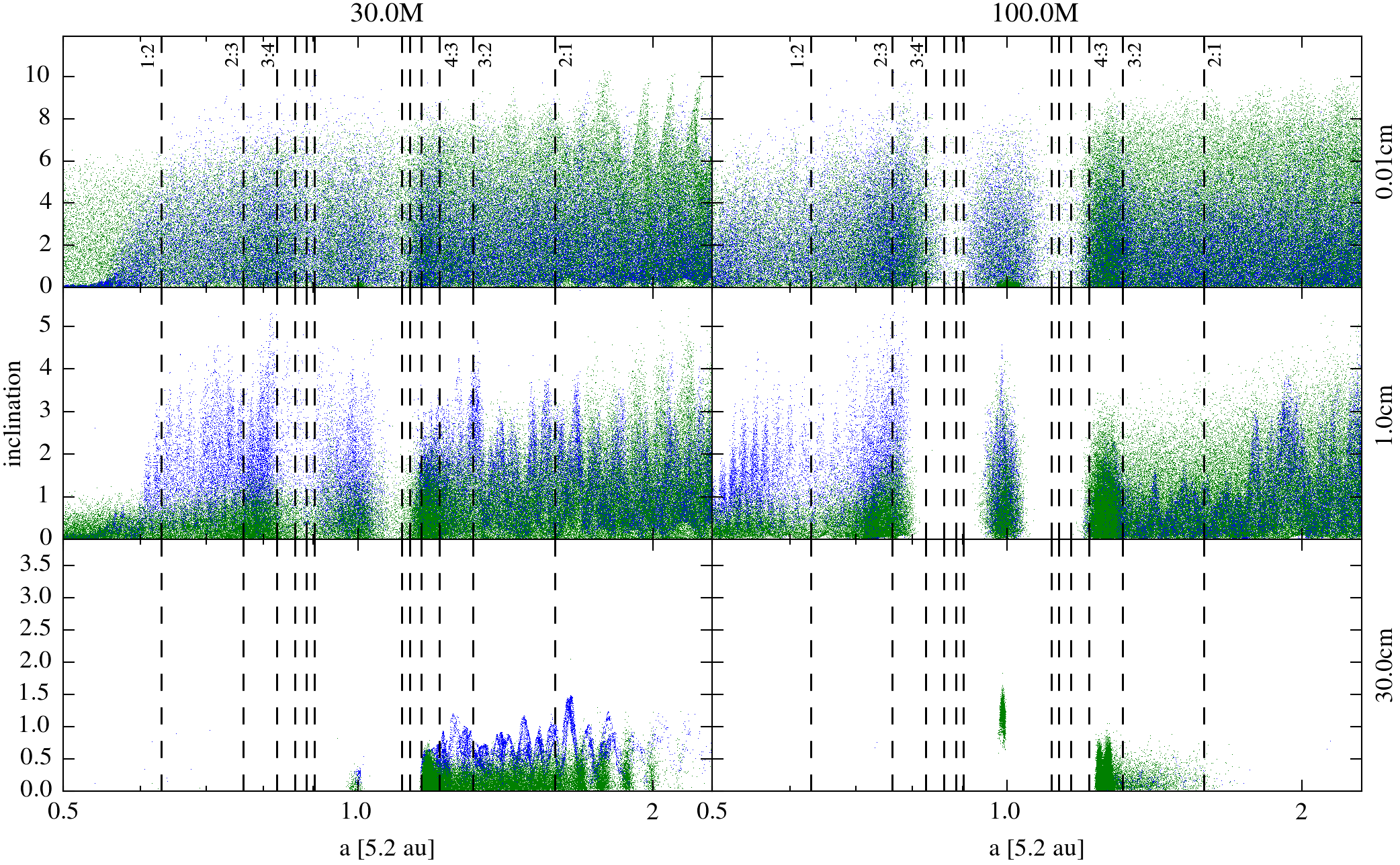}
        \caption{Inclination distribution for $3$ representative size particles in the VSI (blue) and alpha-disc (green) model at the end of the simulation. The first six inner and outer first-order MMRs are overlaid.}
    \label{fig:inc}
\end{figure*}

\subsection{Gap opening}

A planet can open a gap in the dust disc even if no clear gap appears in the gas distribution \citepads[see e.g.][]{2015A&A...584A.110P}. We analysed the radial distribution of the dust population by splitting the computational domain into $400$ logarithmically spaced bins and following its evolution with time.
In Fig.~\ref{fig:histRadial} we plott the distribution at the end of the simulation ($200$ planetary orbits) for the same three representative size particles as in the previous plot. In the left column, the $\unit[10]{cm}$ dust particles (corresponding to a Stokes number of $\tau_\mathrm{s} = 0.08$) do not show a strong perturbation by the presence of the small mass planets (in the first two rows, corresponding to $5$ and \unit[$10$]{M$_\oplus$}). A clear gap appears starting with the \unit[$30$]{M$_\oplus$} planet (third row), where the influence of the VSI (represented by a blue line) favours the formation of deeper gaps.

The intermediate case of meter-sized particles, which corresponds to a Stokes number of $\tau_\mathrm{s} = 1.2$, represents the fastest evolving particles in the simulation; see eq.~(\ref{eq:radialdrift}). As shown in the central column, the distribution is strongly affected by the vertical motion of the VSI where a bunching behaviour can be noticed \citepads{2016A&A...594A..57S}.
This feature cannot be reproduced by the viscous $\alpha$-disc model. In the \unit[$30$]{M$_\oplus$} planet case (third column) we see that the VSI also leads to a faster dust filtration process, which is already completed for the \unit[$100$]{M$_\oplus$} case in which the inner disc is practically devoid of particles.
For the massive planets one notices that a large number of small and large particles remain at the co-rotation location.
These are particles collect near the two Lagrange points L4 and L5.

For a planetesimal-sized object of $\unit[10]{m}$ in the third row (corresponding to a Stokes number of $\tau_\mathrm{s} = 67)$ the gas influence is negligible. A gap is visible in the distribution already for the small mass planets due to their gravitational interaction with the particles. In this case, the VSI does not affect the evolution of planetesimals and yields results identical to the viscous case.

The effect of filtering and gap formation can be understood in terms of the angular velocity distribution in the disc around the planet.
The onset of gap formation leads to a super-Keplerian flow just outside of the planet, which coincides with a maximum in the radial pressure distribution.
The angular velocity is shown in Fig.~\ref{fig:vphiKep}, where the ratio of $v_\phi/v_\mathrm{kep}$ is shown for the four different planetary masses.
Outside of the planet the ratio is slightly smaller than one owing to the pressure support of the gas.
For the planet masses displayed the super-Keplerian motion begins to show for the \unit[$30$]{M$_\oplus$} case.
Hence, as expected the filtering process is directly related directly to the maximum in the angular velocity.
The property that particles (with unit Stokes number) cannot be accreted above a critical planet mass is referred to as the pebble isolation mass.
In appendix \ref{subsec:app-longterm} we present additional simulations to confirm that our simulations
have been run long enough to draw this conclusion about the super-Keplerian motion.

\begin{table*}[tb]
  \caption{Comparison of the mean eccentricity and inclinations of the solids in the disc for the different planet masses at the end of the simulation.}
  \label{tab:incecc}
  \centering
  \begin{tabular}{l | c | c | c | c}
    \hline\hline
          Planet masses & \multicolumn{2}{c|}{Eccentricity} & \multicolumn{2}{c}{Inclination} \\
    \hline
           & VSI [$10^{-3}$] & Laminar [$10^{-3}$] & VSI [$10^{-1}$ deg] & Laminar [$10^{-1}$ deg] \\
    \hline
          $5 M_\oplus$  & 4.673 & 4.480 & 7.439 & 6.920 \\
          $10 M_\oplus$ & 5.054 & 4.786 & 7.310 & 6.936 \\
          $30 M_\oplus$ & 5.062 & 5.250 & 6.071 & 6.702 \\
          $100 M_\oplus$ & 8.843 & 9.406 & 4.854 & 7.089 \\ 
    \hline
  \end{tabular}
\end{table*}

\subsection{Planet-solid disc interaction}

The planet is not only able to open a gap in the dust and gas disc, but it can also generate non-axisymmetric features in their distribution that might be observable with modern observational facilities.
In Fig.~\ref{fig:histTime2} we show the surface density distribution of the dust population after $80$ planetary orbits.
The spiral arms that are typically generated by embedded planets are only (barely) visible 
for the most massive planet (bottom row) and the smallest particles (left column) that are well coupled to the gas dynamics.
For the \unit[$100$]{M$_\oplus$} planet strong vortices are created in the gas disc for the VSI case and less so for the
laminar case \citepads{2017A&A...604A..28S}. Because they are pressure maxima, particles tend to accumulate in these vortices,
which is reflected in the corresponding particle distributions as seen in the bottom left part of the plot.

Also visible in the middle column is the strong effect that the VSI has on pebble-sized particles creating regions where particles are collected as shown in \citetads{2016A&A...594A..57S}. For the large planets we see, as shown in more detail before, in the bottom part of the middle column, that a sort of transition disc is formed since the planet has reached the pebble isolation mass, stopping the influx of meter-sized particles \citepads[as seen also in][]{2012MNRAS.423.1450A}. On the other hand, for the planetesimal-sized objects, shown in the right panel, we observe ripples close to the planet location due to the excitation of the eccentricity in the dust particles by the planet that the gas is not able to effectively damp on a short timescale. Furthermore, the planetesimals are collecting in the Lagrangian points (L4 and L5) in front and behind the planet location, and their density is enhanced at the outer 2:1 mean motion resonance with the planet, visible in the upper part of the last column for the \unit[$100$]{M$_\oplus$} planet.

The eccentricity distribution indicated in Fig.~\ref{fig:ecc} shows only small differences between the $\alpha$- and VSI-disc models, while they are much more pronounced for the inclination distribution shown in Fig.~\ref{fig:inc}.
The excitations at the resonance locations are visible for the biggest size objects in the lower panel. From eq.~(\ref{eq:rminres}) we find that the region where the chaotic behaviour prevents the resonances from stopping planetesimal objects starts at a radial distance of $0.1478$ from the planet location for the \unit[$100$]{M$_\oplus$} planet and $0.1048$ for \unit[$30$]{M$_\oplus$}. These values roughly correspond to the $5$:$4$ and $7$:$6$ MMRs with the planet.
From Fig.~\ref{fig:ecc}, where the location of the major MMRs are plotted, we can confirm this finding.
Planetesimal objects cannot reach the region inside the $5$:$4$ resonance with the \unit[$100$]{M$_\oplus$} planet, while for the \unit[$30$]{M$_\oplus$} planet the bodies are able to reach the $7$:$6$ resonance where the chance of being accreted by the planetary core is much higher.

In Tab.~\ref{tab:incecc} we also report the mean values of all the solids in the disc at the end of the simulation. The eccentricity and inclinations are on average higher for the VSI discs with respect to the laminar discs in the small mass planets. This effect can be explained by the highly anisotropic turbulence nature of the VSI, which is able to stir up dust particles exciting their orbital elements more efficiently. This trend however is inverted for the high mass planets, possibly because the VSI strength is partly reduced by the presence of massive planets, and the production of vortices in which the particles tend to be collected.

The viscous $\alpha$-disc model cannot correctly reproduce the distributions observed in the VSI disc because
we assumed a constant $\alpha$-value throughout the whole computational domain, and did not distinguish between
radial and vertical angular momentum transport. However, as shown recently the VSI turbulence behaves strongly anisotropic 
with a large difference between radial and vertical transport \citepads{2017A&A...599L...6S}.
Since the turbulent kicks in the particle motion are generated based on the constant alpha value,
they over- or under-predict the turbulence efficiency in the laminar disc resulting in a different particle scale height, and thus inclination.
Nevertheless, this does not seem to play a crucial role in the solid accretion rate to the planet from small turbulent velocities.

\begin{figure}[ht]
    \centering
    \includegraphics[width=\columnwidth]{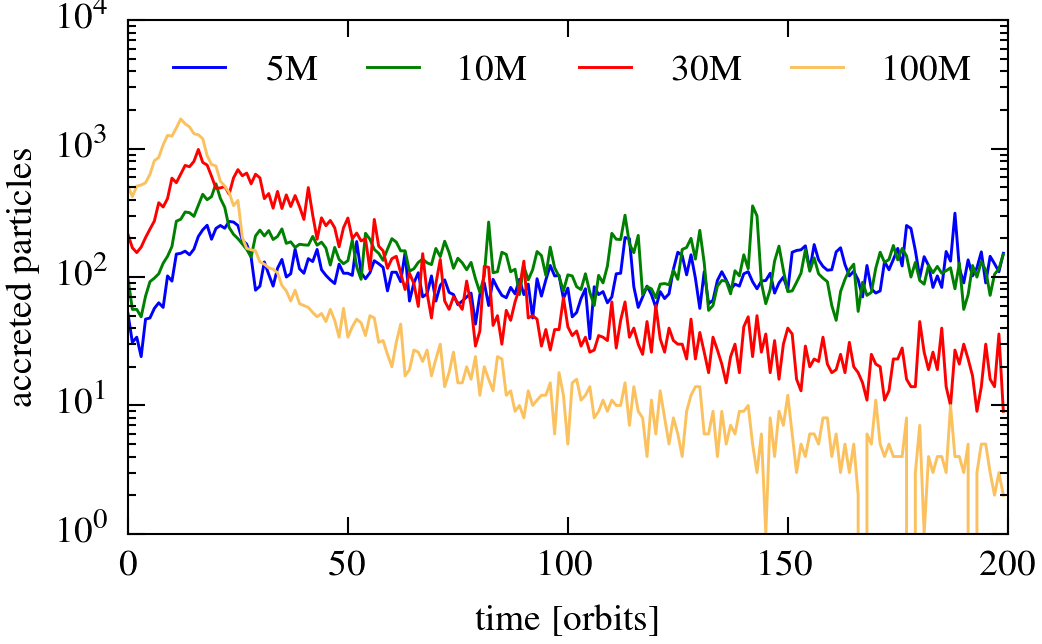}
    \caption{Number of accreted particles per orbit over time for the four different planet masses.
     Shown is the total number summed over all size bins.}
    \label{fig:accretion-evol}
\end{figure}

\section{Solid accretion rate}\label{sec:solidacc}

In order to detect the particles that accrete onto the planet, we adopted two different approaches, depending on the ratio of their Stokes number and the time they spend inside the Hill sphere which, for pebble-sized particles with $\tau_\mathrm{s}$ is $t_\mathrm{enc} = R_\mathrm{H}/\Delta v$, where $\Delta v$ is the relative velocity between the particle and the planet. For particles with stopping time shorter than $t_\mathrm{enc}$, their trajectory close to the planet location is determined primarily by the drag force.
Whether the particle is then accreted depends on the relative strength of the gravitational attraction and the drag force.
If the drag force dominates, it can sweep out a particle even if it is gravitationally bound.
Thus we checked if the timescale for gravitational attraction $t_\mathrm{g} = \Delta v/g$ is shorter than four times the timescale for the stopping time.
The factor of four stems from the results of \citetads{2010A&A...520A..43O}, who had found in their numerical simulations that only a small deflection of a fourth of the velocity is needed to accrete the particle.

Larger particles lose only a small amount of momentum through drag when they cross the Hill sphere. Thus we checked for whether particles inside the Hill sphere are bound by the gravity of the planet, which is the case if the particle has not enough kinetic energy to leave the Hill sphere, that is
\begin{equation}
    e_\mathrm{kin} + e_\mathrm{grav} < e_\mathrm{grav}(R_\mathrm{H})\,.
    \label{eq:accHill}
\end{equation}
Both approaches also agree in the transition region where the stopping time is similar to $t_\mathrm{enc}$.
We checked for these conditions every tenth of a planetary orbit and we flagged as accreted the particles that fulfil the previous criteria without removing these particles from the computational domain.

In Fig.~\ref{fig:accretion-evol} we show the number of accreted particles per orbit for the different planet masses as a function of
time. In the initial phase of the simulations, the number of accreted particles increases while the mass of the
inserted planets grows to their final value (within 20 orbits).
After that, the number of accreted particles drops continuously and
settles roughly to a constant value for the lower mass planets, as there is at least for the faster drifting particles
a continuous supply from the outer disc (see Fig.~\ref{fig:histTime} middle column).
For the larger mass planets, the accretion rates are further reduced
as they have reached their isolation mass for the particles with Stokes number around unity.
Additionally, for the large planets, the small and large particles drift very slowly and, after the particles within the horseshoe region have been accreted, the new inflow from the outer disc is very slow.

\begin{figure}[htbp]
    \centering
    \includegraphics{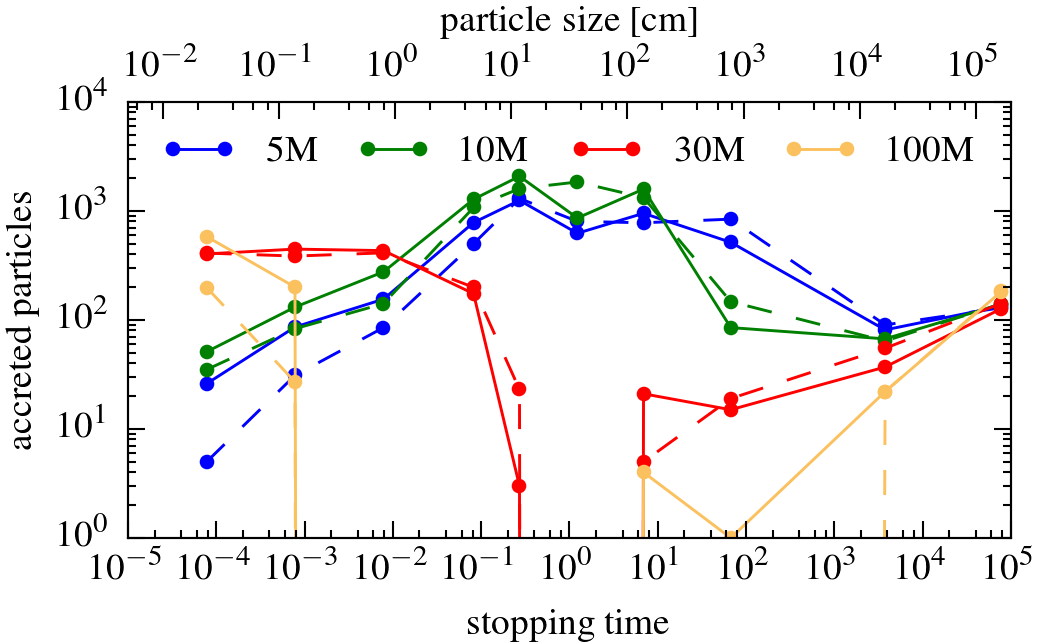}
        \caption{Accreted particles as a function of the particle stopping time, where their approximate size is also reported in the top x-axis, integrated over 50 planetary orbits (from $t=100 - 150$). The VSI (solid line) and alpha disc model (dashed line) are compared.}
    \label{fig:accretion3}
\end{figure}

In Fig.~\ref{fig:accretion3} we show the number of accreted particles of various sizes for different planet masses summed over 50 planetary orbits,
from $t = 100 - 150$.
Several interesting features can be noted. The number of accreted particles peaks for particles in the range between $30$ and \unit[$300$]{cm} (corresponding to pebble-sized objects with Stokes number of order unity) for the small planetary masses, while these particles are effectively accreted and filtered for the two higher mass planets.
The total number of accreted particles adds up to about $5000$ for each of the lower mass planets in agreement with the results shown in Fig.~\ref{fig:accretion-evol}.
Moreover, although the effect of VSI seems marginal it shows in nearly all cases a slightly higher solid accretion rate than for the viscous
$\alpha$-disc model. As the difference is rather small we may conclude that our modelling of the stochastic motion of particles in discs
also gives reasonable results for the accretion rates of the particles onto embedded planets.
To obtain an actual mass accretion rate onto the planet, we need then to convolve this result with a dust size distribution. \citetads{2012A&A...539A.148B} showed that the size distribution is very steep for Stokes numbers less than $0.1$. At that point, there is a gap due to the so-called meter-sized barrier. This effect removes the peak point of our accreted particles for the small mass planets and renders the accretion growth for smaller dust particles even steeper. For the bigger size objects, there are far fewer constraints from models. The leading theory of streaming instability predicts that the peak of formed planetesimals is $10-100$ kilometer-sized objects with a tail in the distribution also to kilometer-sized objects, which represent our bigger size objects in the simulation
\citepads{2016ApJ...822...55S,2017ApJ...847L..12S}.

\begin{figure}[htbp]
    \centering
    \includegraphics[width=0.8\columnwidth]{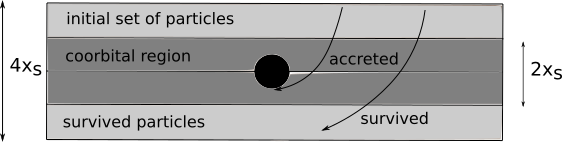}
    \caption{To measure the accretion (and survival) efficiency, the evolution of particles initially within a ring outside the planet co-orbital region (defined as in eq.~\ref{eq:hs_width}) is monitored.
    }
    \label{fig:peff}
\end{figure}

\begin{figure*}[htbp]
    \centering
    \includegraphics{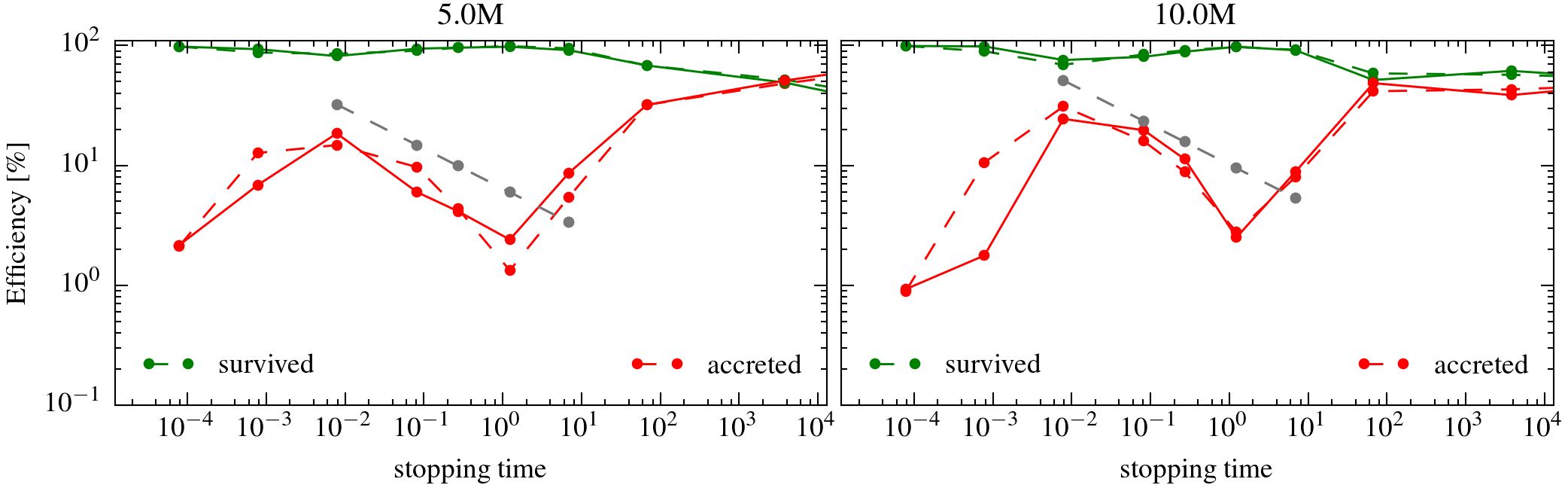}
    \caption{Efficiency of accreted (red line) and survived particles (green line) as a function of Stokes number for different planetary masses. The laminar disc run is represented with solid lines, while the VSI run with dashed lines. The fit from \citetads{2014A&A...572A.107L} is overplotted for intermediate $\tau_\mathrm{s}$ values with a dashed grey line (see eq.~\ref{eq:Lamb14}).
     }
    \label{fig:survivors}
\end{figure*}

\subsection{Efficiency of pebble accretion}

Very important for the mass growth of a planetary core is the efficiency, $P_\mathrm{eff}$, of the accretion process, i.e. the number
of accreted particles onto the planet divided by the number of particles that would otherwise drift across the location of the planet in an
unperturbed disc. Following \citetads{2010A&A...520A..43O}, we define this efficiency as
\beq
\label{eq:Peff}
    P_\mathrm{eff} =  \frac{\dot{M}_\mathrm{acc}}{\dot{M}_\mathrm{drift}}  \,,
\eeq
where $\dot{M}_\mathrm{acc}$ is the actual accretion rate of solids onto the protoplanet and $\dot{M}_\mathrm{drift}$ the particle drift rate through the disc,
given by
\beq
\label{eq:Mdot-drift}
     \dot{M}_\mathrm{drift} =  2 \pi r \Sigma_\mathrm{p} v_\mathrm{drift} \,,
\eeq
where $\Sigma_\mathrm{p}$ is the particle surface density and $v_\mathrm{drift}$ is given by eq.~(\ref{eq:radialdrift}).
The quantity $P_\mathrm{eff}$ in eq.~(\ref{eq:Peff}) is, in fact, the probability that a particle that drifts through the disc is accreted
by the protoplanet.

When analysing the data in this way we encountered the problem that for the particles with very small and very large
Stokes numbers the drift velocities $v_\mathrm{drift}$ are very small (in agreement with eq.~\ref{eq:radialdrift}) such that the calculated
efficiencies became very high because of the relatively large amount of particles accreted.
The reason for this lies in the fact that the accreted small and large particles originate primarily from the horseshoe region and did not migrate to the planet, which is only a transient effect visible in the initial phase of the simulations. Hence, we decided to use an alternative way of measuring the efficiency
of particle accretion from our simulation, which is illustrated in Fig.~\ref{fig:peff}.
To measure the accretion efficiency of particles on to a growing planet we monitor the evolution of particles
that are initially in a ring just outside the planet.
We use the radial range from $x_\mathrm{s}$ to $2 x_\mathrm{s}$, where $x_\mathrm{s}$ is the horseshoe half-width as defined, by
\citetads{2009MNRAS.394.2297P}, as
\beq
\label{eq:hs_width}
       x_\mathrm{s} = 1.68 \, R_\mathrm{p} \,  \left( \frac{q}{h} \right)^{1/2} \,.
\eeq
Radial drift brings the particles into the co-orbital region of the planet. Some of the particles are accreted (and marked so),
while others are able to cross the horseshoe region and are not accreted.
These latter particles enter the inner region of the domain and are called the survivors.
The results of using this procedure for our simulations are shown in Fig.~\ref{fig:survivors} for the different planet masses and
particle sizes.

For the two larger planet masses (\unit[$30$]{M$_\oplus$} and \unit[$100$]{M$_\oplus$}) the results are not very meaningful because the total number of accreted particles is very small as they have already reached their isolation masses. Hence, for the growth of planets, we focus on the two smaller mass planets (\unit[$5$]{M$_\oplus$} and \unit[$10$]{M$_\oplus$}).
In both cases the lowest accretion efficiency is reached for particles with Stokes number $\tau_\mathrm{s} \approx 1$.
For smaller and larger particles the efficiency rises but due to the very slow drift speeds becomes unreliable for very small
($\tau_\mathrm{s} < 10^{-2}$) and large particles ($\tau_\mathrm{s} > 10 - 100$).
The result shows that particles with fast radial drift ($\tau_\mathrm{s}$ around unity) have small accretion efficiencies because a high percentage of particles can cross the planetary orbit.
For Stokes number $\tau_\mathrm{s} =1,$ we find $P_\mathrm{eff} \approx 1.6\%$ for the \unit[$5$]{M$_\oplus$} and around $3\%$ for the \unit[$10$]{M$_\oplus$} planet.

\begin{table}[tb]
    \caption{Comparison of the efficiency of pebble accretion with Stokes number one, for three cases.
    }
    \label{tab:Peff}
    \centering
    \begin{tabular}{l | c | c }
        \hline\hline
        Model &  \multicolumn{2}{c}{Planet Masses}    \\
        \hline
              &   $5 M_\oplus$  &   $10 M_\oplus$    \\
        Hill (eq.~\ref{eq:f_Hill})  & $0.43$  &   $0.50$    \\
        Col (from eq.~\ref{eq:Mdot_col})  &  $0.67$   &   0.80      \\
        turb (simulations) &  $1.6 \cdot 10^{-2}$   &  $3 \cdot 10^{-2}$  \\
      \hline
    \end{tabular}
\end{table}

We can compare our measurements to the results of previous estimates using particle trajectories in the vicinity of the planet.
We use our set-up with the protoplanet located at $R_\mathrm{p} = \unit[5.2]{\mathrm{au}}$.
The radial drift speed is then $v_\mathrm{drift} = \unit[30]{\textrm{m/s}}$ and for the particle density we have with $1\%$ in solids $\Sigma_\mathrm{p} = \unit[2]{\mathrm{g/cm}^2}$.
To calculate the efficiency we use eq.~(37) from \citetads{2010A&A...520A..43O}  with $\log_{10} {P_\mathrm{col}} = 0.5$ for the (dimensionless) collision rate, and drop the 3D correction.
The specific collision rate $P_\mathrm{col}$ is given in this case by (see their eq.~3)
\beq
  \label{eq:Mdot_col}
    \dot{M}_\mathrm{col} = P_\mathrm{col} \Sigma_\mathrm{p} \, R_\mathrm{H} v_\mathrm{H} \,,
\eeq
where 
\beq
  \label{eq:v_Hill}
   v_\mathrm{H} = \Omega_\mathrm{K} R_\mathrm{H}
\eeq
is the Hill velocity.
For an alternative comparison, we use the accretion rate in the Hill regime as given in
\citetads{2012A&A...544A..32L}, also in the 2D version
\beq
  \dot{M}_\mathrm{Hill} = 2 \Sigma_\mathrm{p} R_\mathrm{H} v_\mathrm{H}\,,
  \label{eq:Mdot_Hill}
\eeq
i.e. the ratio of these two is given by $10^{0.5}/2 \approx 1.6$.
To compare directly to other estimates we can transform our obtained accretion rates in terms of the
Hill accretion rate and define
\beq
\label{eq:f_Hill}
    f_\mathrm{Hill}  \equiv  \frac{\dot{M}_\mathrm{acc}}{\dot{M}_\mathrm{Hill}} = P_\mathrm{eff} \, \frac{\dot{M}_\mathrm{drift}}{\dot{M}_\mathrm{Hill}}
         = P_\mathrm{eff} \, \pi \, \frac{R_\mathrm{p} v_\mathrm{drift}}{R_\mathrm{H} v_\mathrm{H}} \,.
\eeq
Applying the definitions of $v_\mathrm{drift}, R_\mathrm{H}$ and $v_\mathrm{H}$ from eqs.~(\ref{eq:radialdrift}, \ref{eq:R_Hill}) and
(\ref{eq:v_Hill}), one obtains with $\taus = 1$ and $\eta = (H/R)^2$
\beq
    f_\mathrm{Hill}  \approx  6.5 \, P_\mathrm{eff} \, \left( \frac{h}{q^{1/3}} \right)^2 \,.
\eeq


Using this equation and our findings for the particle accretion we obtain the results quoted in Table~\ref{tab:Peff}.
We notice that our estimates are about 50\% lower than the 2D approximation for the Hill case, but here we also conside the third dimension,
which can lower the amount of particles within the reach of an embedded planet.
In Fig.~\ref{fig:survivors} we compare also the accretion efficiency with the prescription from \citetads{2014A&A...572A.107L}, that is
\beq
  \label{eq:Lamb14}
  P_\mathrm{eff,LJ} \simeq 0.034 \left(\frac{\tau_\mathrm{s}}{0.1}\right)^{-1/3} \left(\frac{M_\mathrm{c}}{M_\oplus}\right)^{2/3}\left(\frac{r}{10\, \mathrm{au}}\right)^{-1/2}\,,
\eeq
where we see that although our result are slightly lower, the scaling with the Stokes number and planetary masses for the intermediate $\tau_\mathrm{s}$, for which our approach was reliable, is consistent.

From our simulation, we may estimate the mass doubling time for our low mass planets, for which we use \beq
     t_\mathrm{double} = \frac{M_\mathrm{core}}{\dot{M}_\mathrm{acc}} \,.
\eeq
With our results on $P_\mathrm{eff}$ we find $t_\mathrm{double} = \unit[20,000]{\mathrm{yr}}$ for the small mass planets.
The region in the disc that can supply this amount of solid material within the time $t_\mathrm{double}$
extends to roughly \unit[$36$]{au} assuming a constant surface density of solids, $\Sigma_\mathrm{p} = \unit[2]{\mathrm{g/cm}^2}$.

  \begin{figure}[htbp]
      \centering
      \includegraphics[width=\columnwidth]{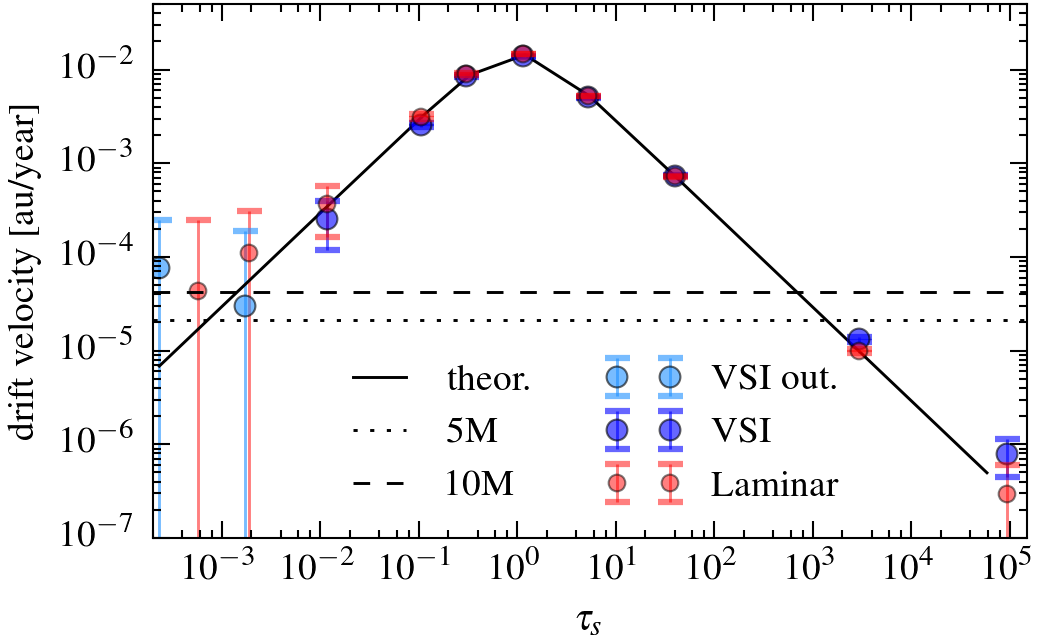}
          \caption{Comparison between the average drift speed of the different dust size particles. These values are calculated at the end of the simulation at $R=1.8$ with the analytical drift speed (black solid line),  the migration speed of a 5\,{M$_\oplus$} (dotted line), and a 10\,{M$_\oplus$} (dashed line) mass planet. The error bars reflect the impact of radial diffusion for the various stopping times. The smaller particle sizes in the VSI simulation are migrating outwards rather then inwards due to the inverse meridional circulation \citetads{2017A&A...599L...6S}. 
      }
      \label{fig:migr}
  \end{figure}

\section{Discussion}\label{sec:disc}

After having presented our main findings we now compare our results to other studies of particles embedded and
accreted onto a planetary core in MHD turbulent discs.
Then we shall discuss possible limitations of our simulations because
in performing the simulations we had to use several approximations to make them feasible.
%
\subsection{Comparison to MHD simulations} \label{sec:compmhd} 

Recently, \citetads{2017ApJ...847...52X} studied the accretion of particles onto small planets embedded in
discs exhibiting magnetically driven turbulence. These authors considered a local shearing box centred at the growing core
and studied three different types of discs: a fully turbulent disc using ideal MHD, a less turbulent disc with ambipolar
diffusion, and a non-turbulent hydrodynamic disc. For all three cases, particles with various Stokes numbers
were injected to the flow after reaching equilibrium, and the accretion rate of particles onto the core was measured.
The measured particle accretion rates were then compared to that in the 2D Hill regime as estimated by \citetads{2012A&A...544A..32L}.
In all cases these authors found for Stokes numbers around unity that their
measured rates agree very well (within $\approx 10\%$) with the 2D accretion rate of eq.~(\ref{eq:Mdot_Hill}) in \citetads{2012A&A...544A..32L}  and argue for a high accretion efficiency.
In our simulations we measure the absolute accretion efficiency as defined in eq.~(\ref{eq:Peff}) and this
is much lower than one for $10^{-2} \leq \tau_s \leq 1$, with an additional drop towards $\tau_s =1$; see Fig.~\ref{fig:survivors}.
To compare directly to \citetads{2017ApJ...847...52X} we can use eq.~(\ref{eq:f_Hill}) and the values quoted in Tab.~\ref{tab:Peff},
which shows that our calculated accretion rates are about 50\% smaller than theirs.

The higher rates obtained by \citetads{2017ApJ...847...52X} may be a result of the relatively small vertical and radial
extent of the disc in their simulations (only one $H$ in each direction) and the fact that they do not consider a stream of
particles through their domain. Hence, they measure the transient accretion rate of particles present initially in the computational box.
In our case we measure the accretion rate of particles in an equilibrium situation. 
The absolute accretion efficiency, as defined in eq.~(\ref{eq:Peff}),  is more important; 
we measure
this rate directly. This is actually very low, but the combination with the large drift speed allows us to tap a
larger reservoir of particles making the accretion of $\taus =1$ particles very useful in the overvall growth process.

\subsection{Planet migration}
  In our simulations, the location of the planet is kept fixed at \unit[5.2]{au}.
  However, planets interact gravitationally with their protoplanetary disc, and this usually results
  in an inwards migration through the disc, which may modify our conclusions about the capture efficiency of particles.
  To check the impact of planet migration we compare the particle drift rates to the expected planet migration for
  which we use the 3D results of \citetads{2010ApJ...724..730D} for low mass planets in the linear regime,
  as stated in \citepads{2012ARA&A..50..211K}
  \begin{equation}
    t_\mathrm{mig} = C \, \frac{M_\star^2}{M_\mathrm{p}\Sigma_g(r_\mathrm{p})r_\mathrm{p}^2} \,
        \left(\frac{H}{r}\right)^2 \, \Omega_\mathrm{K}^{-1}
  ,\end{equation}
  where $C=1/(1.36 + 0.62\beta_\Sigma + 0.43\beta_\mathrm{T})$. The value $\beta_\Sigma = p-1$ 
  is the coefficient of the surface density profile, while $\beta_\mathrm{T} =q$ is the coefficient of the temperature radial profile.
  In our models we used  $\beta_\Sigma = 0.5$ and $\beta_\mathrm{T} = 1.0$ (see Sect.~\ref{par:gasdisc}) and then we find $C = 0.48$.
  In Figure~\ref{fig:migr} we compare the dust drift velocity measured at the beginning of the simulations, when the planet has not yet perturbed the disc structure, to the theoretical drift speed (as obtained from eq.~\ref{eq:radialdrift}; black solid line),
  and the migration speed of the two small planets (two dashed lines).  We can see that only the tails of the particle size distribution drift slower than the planets.
  However, pebble-sized particles have a migration speed orders of magnitude faster than the planet.
  Hence, we conclude that any planet migration does not influence our results for pebble accretion.
  For larger mass planets that migrate with Type II migration, the drift speed slows down considerably, and their
  masses are well above the pebble isolation mass. The impact of non-circular planetary orbits on the pebble accretion
 efficiency was studied recently by \citetads{2018arXiv180306149L} who found that it can be increased slightly for moderately 
 eccentric orbits.

\subsection{Equation of state}
  In our simulations, we used an isothermal equation of state and now briefly discuss a possible impact of
  including radiative effects. The inclusion of radiative transfer leads to finite cooling times of the gas that lowers the efficiency of the VSI-driven turbulence \citep{2013MNRAS.435.2610N}.
  In full simulations that include radiative transfer it has been shown that in irradiated discs an efficiency of
  $\alpha \approx 10^{-4}$ can be reached \citep{2014A&A...572A..77S, 2016A&A...594A..57S}, while \citet{2017ApJ...850..131F} find
  a somewhat smaller value. All those simulations apply to larger distances from the star and it remains to be seen what the
  VSI-efficiency is at shorter distances from the star. In any case, a reduced turbulence level leads to a concentration
  of the dust particles in the midplane, which might enhance the accretion process.
  On the other hand, the inclusion of radiative transfer allows for additional disc heating by the planet
  (by the spiral waves), which enhances the disc temperature and might lead to partial evaporation of the particles.
  However, for the lower mass planets, for which the dust accretion efficiency is higher, the effect on the disc is not be that strong
 and we do not expect a large impact. Additionally, the dust clearing around the planet alters the opacity
 of the medium and hence the radiative transport. These impacts of radiative transport and the link to observations
 have to be investigated in more detail by future simulations.
\subsection{Dust feedback}
  In our simulations we have neglected the backreaction of the dust onto the gas. Within a disc without an embedded planet
  the particle concentrations are such that the dust density remains typically smaller than the gas density, given an initial dust to
  gas ration of 1/100.
  In the presence of a planet, this is not true in the case of filtering because then the dust density can equal the gas
  density near the pressure maximum. This situation has recently been explored
  by \citetads{2018arXiv180107971W}, who showed that dust feedback can potentially displace the gas density maximum,
  and thus the pressure maximum, outwards.
  Additional dust diffusion (for example from disc turbulence) can smooth the density peak of the dust distribution,
  altering the dust filtration process for particles with Stokes numbers around unity.
  However, concerning the filtration ability, which also affects pebble accretion onto the planet,
  they did not observe any difference by adding dust feedback.
  Hence, we conclude that dust feedback does not impact our results significantly.
  In a realistic scenario, where a dust size distribution is present, this effect is even less pronounced.
  The impact of dust feedback onto the dust dynamics in the dust trap itself will have to be investigated in more detail in the future.
\subsection{Numerical convergence}
  Our hydrodynamical simulations are performed with one numerical resolution as given in Table~\ref{Tab:sum}.
  This is based on our results in \citetads{2017A&A...604A..28S} in which we studied the effect of doubling the grid resolution on the VSI
  and found no noticeable differences in the disc dynamics.
  The calculated $\alpha_\mathrm{SS}$ close to the inner boundary (see their Figure~1) was increased marginally,
  however it had no effect at the location of the planet.
  We also checked in \citetads{2017A&A...604A..28S} that the torque acting onto the planet reached a constant value (see their Figure~6),
  guaranteeing that the model was run long enough for the disc-planet system to have reached a quasi-stable state. The obtained
  torque distributions on the planet were also identical for the standard resolution (used here) and the simulations with doubled
  resolution.
  The dust dynamics is not affected by numerical resolution since we did not take into account their backreaction on the gas. 
  Thus, we do not expect that our results on accretion efficiencies and dust dynamics in the vicinity of the planet are
  impacted by resolution that is too low.

\section{Conclusions}\label{sec:concl}
In this study we have modelled the dynamics of a broad range of solid particles, ranging from \unit[$100$]{$\mu$m} dust particles to kilometer-sized planetesimals, interacting with a growing planetary core in a 3D globally isothermal disc. By modelling a global disc, we were able to take into account the effect of MMRs for planetesimal-sized objects and the enhanced particle density in spiral arms and vortices. The turbulence driving the disc evolution has been modelled both self-consistently through the VSI instability and with an alpha parameter derived from the VSI simulation where the turbulence has been recreated in the particle dynamics by adding random kicks to their motion. We determined the solid accretion rate onto the planet after it reaches a stable state averaging the values over \unit[$50$]{planetary orbits} in Fig.~\ref{fig:accretion3}.

The actual growth rates in particles that a planet can achieve depends on the particle size distribution.
In our study, we sample the particle dynamics in ten different size bins. One can convolve this result with a model of dust size distribution to obtain a mass accretion rate onto the planet.
We observed a peak in the absolute number of accreted particles in the range of pebble-sized objects (100 cm) with Stokes number of order unity,
but the strength of this effect depends strongly on the chosen dust size distribution.
Concerning the accretion efficiency we find that the minimum efficiency is reached for particles with $\tau_\mathrm{s}=1,$ where $P_\mathrm{eff} = 0.016$ and $0.03$ for the planets with masses \unit[$5$]{M$_\oplus$} and \unit[$10$]{M$_\oplus$}, respectively.
For smaller and larger particles the efficiency rises but due to the rapid inwards drift of particles with $\tau_\mathrm{s} =1$, we find that the optimal particle size for pebble accretion for our massive cores is about one metre at the orbit distance of about \unit[$5$]{au}.
If all the solid material in the disc was this size range, the mass doubling time would be around 20,000 yrs.
We find that the obtained accretion efficiencies are very similar for the VSI turbulent disc and the laminar disc models,
one has to keep in mind however to add the stochastic kicks to the particles for the viscous model. This similarity
can be attributed to the fact that the overall turbulence generated by the VSI is relatively weak such that the disc structures
are very similar, despite the occurrence of vortices in the VSI case.

The accretion efficiency found in our simulations agrees reasonably well with previous results, for example, the 2D approximations of \citetads{2010A&A...520A..43O} and \citetads{2012A&A...544A..32L} or the 3D turbulent simulations of \citetads{2017ApJ...847...52X} who found similar results for particles with $\tau_\mathrm{s}=1$.
To obtain the efficiencies of very small or large particles exactly one needs longer integration time due to the very slow radial drift.
Concerning the pebble isolation mass of a growing planet we confirm that the occurrence of a pressure maximum in the
gas created by the planet is sufficient to filter particles with Stokes numbers of unity efficiently, at least for the relatively weak
VSI turbulence. Hence, using purely hydrodynamical studies the dependence of the isolation mass on the viscosity and pressure
scale height of the disc has been examined recently to obtain scaling relations \citepads{2018arXiv180102341B}. 

The treatment of the turbulence adopted for the particles in the laminar disc produced accretion rates in good agreement with those of the self-consistent VSI treatment. The impact of radiative transfer within the disc and the migration of the planet through the disc were not treated. These topics need to be addressed in future work.

\begin{acknowledgements}
    We thank the anonymous referees for useful comments and suggestions.
    The very helpful discussions with Chris Ormel are gratefully acknowledged. 
    G. Picogna acknowledges the support through the German Research Foundation
    (DFG) grant KL 650/21 within the collaborative research programme ``The first
    10 Million Years of the Solar System''.
    M.H.R. Stoll acknowledges the support through the (DFG) grant KL 650/16.
    This work was performed on the computational resource ForHLR I funded by the Ministry of Science,
    Research and the Arts of Baden-W\"urttemberg, and the DFG.
    This research was supported by the Munich Institute for Astro- and Particle Physics
    (MIAPP) of the DFG cluster of excellence ''Origin and Structure of the Universe''.
\end{acknowledgements}

\bibliographystyle{aa}
\bibliography{calibre,biblio}{}

 \onecolumn
\begin{appendix}

  \section{Long-term two-dimensional integrations}\label{sec:longterm}
\label{subsec:app-longterm}
  \begin{figure}[htbp]
      \centering
      \includegraphics[width=0.9\columnwidth]{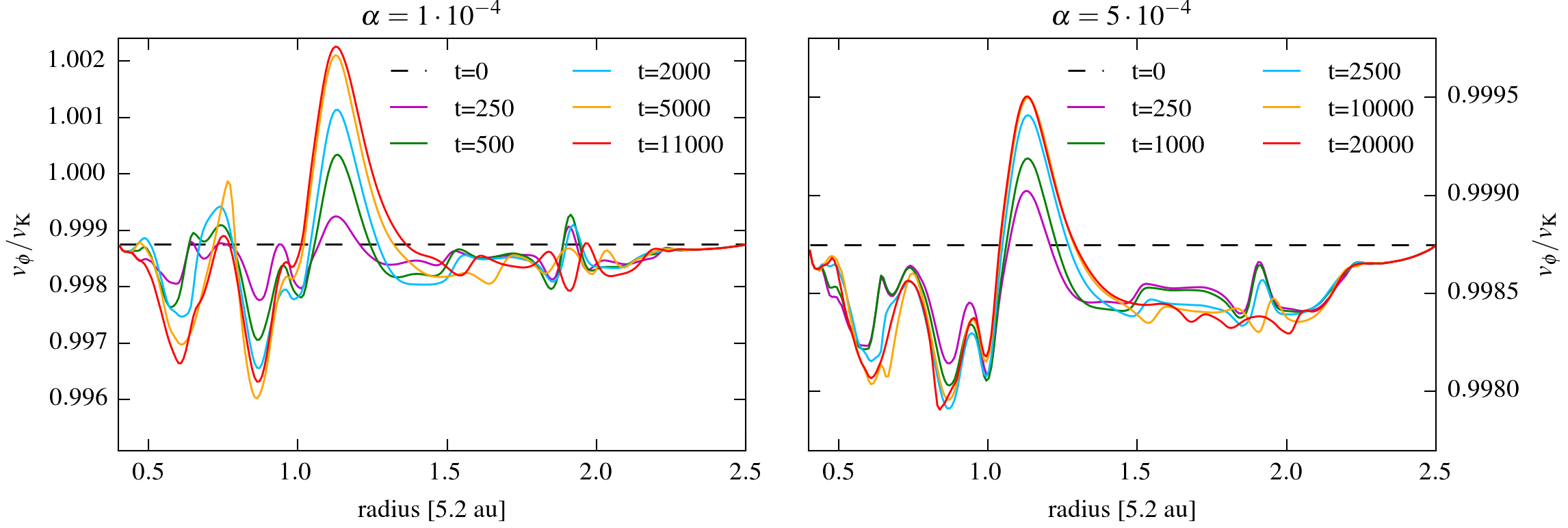}
      \caption{Ratio of the azimuthal gas velocity to the Keplerian value for a 2D
    disc with an embedded 10 $M_\oplus$ planet at different times after insertion of the planet. 
   The simulation in the left panel uses a viscosity of $\alpha = 10^{-4}$ and the right  $\alpha = 5 \times 10^{-4}$.
      \label{fig:omega2D}
      }
  \end{figure}

  In Figure \ref{fig:vphiKep} above we displayed the ratio of the angular velocity of the gas to the Keplerian velocity
  for the different planet masses and compared the turbulent case to the viscous laminar case.
  From this, we argued that the occurrence of super-Keplerian flow can be taken as an indication of having reached the
  isolation mass for that particular planet. However, this argument is only valid if the equilibrium of the flow has already been
  reached and does not change in time significantly anymore. 
  For a viscous disc with a kinematic viscosity $\nu$, the viscous timescale
  is given by
 \beq
     \tau_\nu = \frac{\Delta r^2}{\nu}  \,,
 \eeq
  where $\Delta r^2$ is the spatial region under consideration.
  Assuming an $\alpha$-type viscosity with $\nu \sim \alpha \Omega_\mathrm{K} H^2$ and $\Delta r = f_\mathrm{H} H$
  one finds for the viscous timescale
 \beq
     \tau_\nu = \frac{f_\mathrm{H}^2}{2 \pi \alpha} \, P_\mathrm{K} \,, \label{eq:tau-nu}
 \eeq
  where $P_\mathrm{K}$ is the Keplerian period. The maximum of $\Omega$ occurs roughly at a distance of $\Delta r = 2 H$
 in our case and for $f_\mathrm{H} = 2$ eq.~(\ref{eq:tau-nu}) gives an equilibration time of over 1200 orbits.
  To run our computations in full 3D for such a long timescale would have been too costly and we investigated this issue
  by performing comparison 2D simulations of planets embedded in flat discs. 
  For these, we used a 10 $M_\oplus$ planet and two different effective viscosities, $\alpha = 5 \times 10^{-4}$
  and a lower viscosity case using $\alpha = 10^{-4}$. The results are shown in Fig.~\ref{fig:omega2D}.
  While for the low viscosity case the flow becomes super-Keplerian, the model with $\alpha = 5 \times 10^{-4}$ 
  remains sub-Keplerian throughout. From this, we infer that in VSI turbulent discs with an effective 
  $\alpha = 5 \times 10^{-4}$ the isolation mass is indeed above 10  $M_\oplus$ as found in the full 3D simulations
  presented above. 
   
  \section{Integrator}\label{sec::integrators}
  We used two different integrators to evolve the Lagrangian particles, based on their coupling with the gas dynamics.

    \subsection{Semi-implicit integrator in polar coordinates}
    The dynamics of particles well coupled to the gas, which have a stopping time much smaller than the time step adopted to evolve the gas dynamics, is described by adopting the semi-implicit Leapfrog (Drift-Kick-Drift) integrator described in~\citetads{2014ApJ...785..122Z} in polar coordinates.
    This method guarantees the conservation of the physical quantities for the long-term simulations performed in this paper and, at the same time, it is faster than an explicit method.

\noindent
    The variables are updated beginning with a first half drift
    \begin{align*}
      L_{\mathrm{r},n+1} & = L_{\mathrm{r},n} \,, \quad \quad 
      &  L_{\theta,n+1} & = L_{\theta,n} \,, \quad  \quad  
      &  L_{\phi,n+1} & = L_{\phi,n} \,.  \\
      r_{n+1} &= r_n + L_{\mathrm{r},n}\frac{\mathrm{d}t}{2} \,, \quad \quad 
      & \theta_{n+1} &= \theta_{n}+ \frac{1}{2}\left(
      \frac{L_{\theta,n}}{r_{n}^2} +
      \frac{L_{\theta,n+1}}{r_{n+1}^2}
      \right) \frac{\mathrm{d}t}{2} \,, \quad \quad 
      & \phi_{n+1} &= \phi_{n}+\frac{1}{2}\left(
      \frac{L_{\phi,n}}{R_n^2}+\frac{L_{\phi,n+1}}
      {R_{n+1}^2}\right) \frac{\mathrm{d}t}{2}\,,
    \end{align*}

\noindent
 followed by a kick step
    \begin{align*}
      r_{n+2} & = r_{n+1}  \,, \quad  \quad 
       \theta_{n+2} = \theta_{n+1}  \,, \quad \quad
      \phi_{n+2} = \phi_{n+1}  \,. \\ 
      L_{\phi,n+2} &= L_{\phi,n+1} +
      \frac{\mathrm{d}t}{1+\frac{\mathrm{d}t}{2t_{\mathrm{s},n+1}}}
      \left[-{\left(\frac{\partial\Phi}{\partial\phi}\right)}_{n+1} +
      \frac{L_{\phi,\mathrm{g},n+1}-L_{\phi,n+1}}{t_{\mathrm{s},n+1}}
      \right] \,, \\
      L_{\theta,n+2} &= L_{\theta,n+1} +
      \frac{\mathrm{d}t}{1+\frac{\mathrm{d}t}{2t_{\mathrm{s},n+1}}}
      \left[\frac{1}{2}\frac{\cos(\theta_{n+2})}{\sin(\theta_{n+2})}\left(\left(\frac{L_{\phi,n+1}}{R_{n+1}}\right)^2+
      \left(\frac{L_{\phi,n+2}}{R_{n+2}}\right)^2\right)-{\left(\frac{\partial\Phi}{\partial\theta}\right)}_{n+1} +
      \frac{L_{\theta,\mathrm{g},n+1}-L_{\theta,n+1}}{t_{\mathrm{s},n+1}}+
      \right] \,,  \\
      L_{\mathrm{r},n+2} &= L_{\mathrm{r},n+1} +
      \frac{\mathrm{d}t}{1+\frac{\mathrm{d}t}{2 t_{\mathrm{s},n+1}}}
      \left[
      \frac{1}{2r_{n+2}}\left(\left(\frac{L_{\phi,n+1}}{R_{n+1}}\right)^2+
      \left(\frac{L_{\phi,n+2}}{R_{n+2}}\right)^2 +
      \left(\frac{L_{\theta,n+1}}{r_{n+1}}\right)^2 +
      \left(\frac{L_{\theta,n+2}}{r_{n+2}}\right)^2\right)
      -{\left(\frac{\partial\Phi}{\partial r}\right)}_{n+1} +
      \frac{L_{\mathrm{r,g},n+1}-L_{\mathrm{r},n+1}}{t_{\mathrm{s},n+1}}
      \right] \,,
    \end{align*}

\noindent
    and, for the laminar disc case, also a random kick, i.e.
    \begin{align*}
      r_{n+2} &= r_{n+2} + \delta r_\mathrm{d,T} \,, \quad \quad
     & \theta_{n+2} &= \theta_{n+2} + \delta \theta_\mathrm{d,T}  \,, \quad \quad
     &  \phi_{n+2} &= \phi_{n+2} + \delta \phi_\mathrm{d,T} \,. \\
    \end{align*}

\noindent
    Finally, a second half drift follows as the first half drift.

  \subsection{Fully-implicit integrator in polar coordinates}
  For particles with stopping time much smaller than the numerical time
  step, the drag term can dominate the gravitational force term, causing the
  numerical instability of the integrator.
  Thus, it is necessary to adopt a fully implicit integrator
  following~\citetads{2010ApJS..190..297B,2014ApJ...785..122Z}.\\

\noindent
    We begin with a predictor step for the particle positions
    \begin{align*}
      r' &= r_n + L_{\mathrm{r},n} \mathrm{d}t \,, \quad \quad
     & \theta' &= \theta_n + \frac{L_{\theta,n}}{r_n^2} \mathrm{d}t \,, \quad \quad
     & \phi' &= \phi_n + \frac{L_{\phi,n}}{R_n^2} \mathrm{d}t \,.
    \end{align*}

\noindent
    followed by a shift for the momenta
    \begin{align}
      L_{\phi,n+1} &= L_{\phi,n} +
      \frac{{\mathrm{d}t}/2}
      {1+\mathrm{d}t\left(\frac{1}{2t_\mathrm{s,n}}+\frac{1}
      {2t_\mathrm{s,n+1}}+
      \frac{\mathrm{d}t}{2t_\mathrm{s,n}t_\mathrm{s,n+1}}\right)}\cdot
      \Bigg[
        -{\left(\frac{\partial\Phi}{\partial \phi}\right)}_n
        -\frac{L_{\phi,n} - L_{\phi,\mathrm{g},n}}{t_{\mathrm{s},n}} + \nonumber \\
        &+ \Bigg(
          -{\left(\frac{\partial\Phi}{\partial \phi}
          \right)}_{n+1}
          -\frac{L_{\phi,n} - L_{\phi,\mathrm{g},n+1}}
          {t_{\mathrm{s},n+1}} \Bigg)
        {\left(1+\frac{\mathrm{d}t}{t_{\mathrm{s},n}}\right)}
      \Bigg] \,, \nonumber \\
      L_{\theta,n+1} &= L_{\theta,n} + \frac{{\mathrm{d}t}/2}{1+\mathrm{d}t \left(\frac{1}{2t_\mathrm{s,n}}+\frac{1}{2t_\mathrm{s,n+1}}+
      \frac{\mathrm{d}t}{2t_\mathrm{s,n}t_\mathrm{s,n+1}}\right)}\cdot
      \Bigg[
        -{\left(\frac{\partial\Phi}{\partial \theta}\right)}_n
        -\frac{L_{\theta,n} - L_{\theta,\mathrm{g},n}}{t_{\mathrm{s},n}} + \frac{\cos(\theta')}{\sin(\theta')}\left(\frac{L_{\phi,n}}{R}\right)^2 \nonumber \\
        &+ \Bigg(
          -{\left(\frac{\partial\Phi}{\partial \theta}
          \right)}_{n+1}
          -\frac{L_{\theta,n} - L_{\theta,\mathrm{g},n+1}}
          {t_{\mathrm{s},n+1}} + \frac{\cos(\theta')}{\sin(\theta')}\left(\frac{L_{\phi}'}{R'}\right)^2
        \Bigg)
        {\left(1+\frac{\mathrm{d}t}{t_{\mathrm{s},n}}\right)}
      \Bigg] \,, \nonumber \\
      L_{\mathrm{r},n+1} &= L_{\mathrm{r},n} +
      \frac{{\mathrm{d}t}/2}
      {1+\mathrm{d}t{\left(
        \frac{1}{2t_{\mathrm{s},n}} +
       \frac{1}{2t_{\mathrm{s},n+1}} +
        \frac{\mathrm{d}t}{2t_{\mathrm{s},n}t_{\mathrm{s},n+1}}
      \right)}}\cdot \Bigg[
        -{\left(\frac{\partial\Phi}{\partial r}\right)}_\mathrm{n}
        -\frac{L_{\mathrm{r},n}-L_{\mathrm{r,g},n}}{t_{\mathrm{s},n}}
        +\frac{1}{r_n}\left(\frac{L_{\phi,n}^2}{R_n^2} + \frac{L_{\theta,n}^2}{r_n^2}\right) + \nonumber \\
        &+ \Bigg(
          -{\left(\frac{\partial\Phi}{\partial r}\right)}_\mathrm{n+1}
          -\frac{L_{\mathrm{r},n}-
          L_{\mathrm{r,g},n+1}}{t_{\mathrm{s},n+1}}
          +\frac{1}{r'}\left(\frac{L_{\phi,n+1}'^2}{R'^2}+\frac{L_{\theta,n+1}^2}{r'^2}\right)
        \Bigg)
        {\left(1+\frac{\mathrm{d}t}{t_{\mathrm{s},n}}\right)}
      \Bigg] \,, \nonumber \\  \nonumber 
    \end{align}

\noindent
    a turbulent kick for the laminar disc case
    \begin{align*}
      r_{n} &= r_{n} + \delta r_\mathrm{d,T} \,, \quad \quad 
     &  \theta_{n} &= \theta_{n} + \delta \theta_\mathrm{d,T} \,, \quad \quad 
     &  \phi_{n} &= \phi_{n} + \delta \phi_\mathrm{d,T} \,, \\
    \end{align*}

\noindent
    and finally a corrector step for the particle positions.
    \begin{align*}
      r_{n+1} &= r_{n} + \frac{1}{2}(L_{\mathrm{r},n}+
      L_{\mathrm{r},n+1})\mathrm{d}t \,,  \quad \quad
      & \theta_{n+1} &= \theta_{n} + \frac{1}{2}\left(
      \frac{L_{\theta,n}}{r_{n}^2}+\frac{L_{\theta,n+1}}
      {r_{n+1}^2}\right) \mathrm{d}t  \,,  \quad \quad
     &  \phi_{n+1} &= \phi_{n} + \frac{1}{2}\left(
      \frac{L_{\phi,n}}{R_{n}^2}+\frac{L_{\phi,n+1}}
      {R_{n+1}^2}\right) \mathrm{d}t \,.
    \end{align*}

\end{appendix}

\end{document}